\input epsf
\documentclass[nohyper,notoc]{article} 
\usepackage{axodraw}
\usepackage{epsfig}

\def\bit{\begin{itemize}}
\def\eit{\end{itemize}}
\def\ben{\begin{enumerate}}
\def\een{\end{enumerate}}
\def\beq{\begin{equation}}
\def\eeq{\end{equation}}
\def\bea{\begin{eqnarray}}
\def\eea{\end{eqnarray}}
\def\bq{\begin{quote}}
\def\eq{\end{quote}}
\def \lsim{\mathrel{\vcenter
     {\hbox{$<$}\nointerlineskip\hbox{$\sim$}}}}
\def \gsim{\mathrel{\vcenter
     {\hbox{$>$}\nointerlineskip\hbox{$\sim$}}}}
\def\gappeq{\mathrel{\rlap {\raise.5ex\hbox{$>$}}
{\lower.5ex\hbox{$\sim$}}}}
\def\lappeq{\mathrel{\rlap{\raise.5ex\hbox{$<$}}
{\lower.5ex\hbox{$\sim$}}}}

\def\checkmark{\surd}

\def\SX{{\bf S}_{L.R}}
\def\SL{{\bf S}_{L}}
\def\SR{{\bf S}_{R}}
\def\T{{\bf T}}

\def\dslash{ \, \partial  \! \! \! \! / ~ }

\def\mec{\mu e ~{\rm conversion}}
\def\meee{\mu \to \bar{e}e \bar{e}}
\def\meg{\mu \to e \gamma}
\def\teg{\tau \to e \gamma}
\def\tmg{\tau \to \mu \gamma}
\def\teLg{\tau \to e_L \gamma}
\def\tmLg{\tau \to \mu_L \gamma}
\def\teRg{\tau \to e_R \gamma}
\def\tmRg{\tau \to \mu_R \gamma}
\def\tlg{\tau \to e_\b \gamma}
\def\llg{e_\a \to e_\b \gamma}
\def\m3e{\mu \to e \bar{e} e}
\def\a{\alpha}
\def\b{\beta}

\def\m{\mu}

\begin{document}

\renewcommand{\thefootnote}{\fnsymbol{footnote}}
\begin{center}
{\Large {\bf Learning about flavour structure from $\tlg$ and
$\meg$?}}

$~$\\

{\bf   Sacha Davidson $^{1,}$\footnote{E-mail address:
s.davidson@ipnl.in2p3.fr} 
}
 
\vskip 10pt  
$^1${\it IPNL, Universit\'e de Lyon, Universit\'e Lyon 1, 
CNRS/IN2P3, 4 rue E. Fermi 69622 Villeurbanne cedex, France
}\\
\vskip 20pt
{\bf Abstract}
\end{center}

\begin{quotation}
  {\noindent\small

Current and upcoming experiments should improve the 
sensitivity to $\llg$ decays by an order of magnitude.
This paper assumes that one of the
$\tlg$ decays is observed, and explores
the structure and consequences of the
required new flavoured couplings. 
In simple models (a low-scale seesaw, leptoquarks)
it is shown that the dipole vertex function
is proportional to a product of 
flavoured matrices from the Lagrangian
(a ``Jarlskog-like'' invariant),
provided that the loop particles are
weakly coupled to the Higgs.  Secondly,
if the dipole vertex function
has a hierarchical structure, 
this can  imply that only some
of the $\tlg$ modes can be observed,
due to the ``approximate zero'' implied
by the bound on $\meg$.  The assumptions underlying
this potential test of
a hierarchical structure are discussed.

\vskip 10pt
\noindent
}

\end{quotation}

\vskip 20pt

\section{Introduction}
\label{intro}

Measuring $\llg$ would indicate New Physics 
near  the TeV scale.
Various studies 
of TeV New Physics  models  anticipate
correlations  in the Lepton Flavour Violating  rates for 
$\llg$ and other processes. These predictions  depend on the particle
content and masses, the flavoured  couplings
of the model,  and
on the parameter space scan. This paper
focuses on  the flavour structure of New Physics:
what can   $\llg$ tell us about patterns
of lepton flavour violation, with limited knowledge  of the
new particle content?
  
After electroweak symmetry
breaking,
the  dipole operator which induces $\llg$,
 can be written, 
\beq
\frac{e}{16 \pi^2} {\Big (}
 [{ X}_{L}]_{ \b \a} \overline{e_\b} \sigma ^{\mu \nu} P_L e_\a F_{\mu \nu}
+
 [{ X}_{R}]_{\b \a} \overline{e_\b} \sigma ^{\mu \nu} P_R e_\a F_{\mu \nu} 
 {\Big )}
~~~.
\label{dipole}
\eeq
where 
$\sigma ^{\mu \nu} = i [\gamma^\mu, \gamma^\nu]/2$.
The dipole ``vertex function''\footnote{The
${\bf X}_{L,R}$ are  arbitrary functions
of coupling constants and masses from
the Standard Model and beyond. They are refered to 
as vertex functions, to distinguish them from
coefficients of operators in an expansion
in $1/\Lambda_{NP}$, where $\Lambda_{NP}$ is
a New Physics scale.}
${\bf X}_{L,R}$  can be expressed as  
a dimensionful unflavoured constant,
multiplied by  a dimensionless
``flavour tensor''.
A first question,
is  when and how  does
the $\llg$  flavour tensor
reflect the flavoured couplings of the
Lagrangian (``spurions'')? We
study, at one loop in  two  models --- the type I
seesaw and  leptoquarks  ---
when the $\llg$ flavour tensor 
can be constructed,
like a Jarlskog invariant \cite{Jinvar}, by contracting
flavoured tensors present in the Lagrangian.

Experimental searches for  Lepton
Flavour Violating (LFV) radiative decays
impose significant bounds on the flavour changing
dipole vertex function. The MEG \cite{MEG} experiment
constrains $BR(\meg) < 2.4 \times 10^{-12}$ and  plans
to reach a sensitivity of  $10^{-13}$
in the coming years. Babar and Belle
have searched for $\tlg$ decays,
and impose  
 $BR(\tmg) < 4.4  \times 10^{-8}$\cite{Babartmgteg,Belletmg}
and
 $BR(\teg) < 3.3 \times 10^{-8}$ \cite{Babartmgteg}
(see
 table \ref{tab}). Super-B factories
expect a
sensitivity $BR \sim 10^{-9}$ \cite{SB,SBatVal,CDRSB}. 
A second aim of  this paper, 
is  to explore whether Super-B factories,
combined with MEG,  can test the flavour
structure of the dipole vertex function.
So we make the non-trivial assumption that
one of the $\tlg$ rates is within the
reach of the Super-B factories. 
 This is
a phenomenologically  interesting
scenario to envisage, because observing one
$\tlg$ rate  gives  two
pieces of information, the
second one being  an ``approximate zero'' in
the dipole flavour tensor,
 imposed by the $\meg$ bound.
We arge that, 
if the flavour
tensor   is hierarchical\footnote{meaning that it is dominated
by its largest eigenvalue, analogously  to
$
[Y_u Y_u^\dagger]_{ij} \simeq y_t^2 V_{ti}^* V_{tj}
$} and certain other
assumptions are satisfied, then 
one of the  remaining $\tlg$
rates is  suppressed  below Super-B sensibilities.
The assumptions required for this potential ``test''
of the hierarchical structure  are discussed.

Section \ref{sec:notn}
  gives
the $\llg$ branching ratios,
 reviews   various properties of
the dipole vertex function  ${\bf X}_{L,R}$, and  presents one-loop
formulae for it. At the end of the section,
is discussed the relation of  ${\bf X}_{L,R}$  to
the  coefficient of the
 dimension six, electroweak gauge
invariant,  dipole operator. 
 In the
models considered in  sections \ref{sec:invar}, 
which are  a TeV-scale
seesaw, and various scalar leptoquarks,
the vertex function  ${\bf X}_{L,R}$
turns out to be
  proportional to an ``invariant'',  when
it is the coefficient of this  dimension six
operator. 
 The definition of what
we mean by ``invariant'', can be
found at the beginning of section \ref{sec:invar}.
Section \ref{sec:hier} reviews the argument
that a hierarchical structure can be tested. 
Finally, the prospects of obtaining
invariants and testing a hierarchy are discussed  in 
section \ref{sec:disc}. 
A summary of useful loop  results  from a paper by
Lavoura\cite{lavoura} appears in the Appendix.

\section{Notation and review}
\label{sec:notn}

For reviews of the expectations and prospects for
lepton flavour violating processes, see
 {\it e.g.} \cite{SBatVal,CDRSB,KunoOkada,ybook}.
The theoretical predictions have been widely studied 
in popular models such as  Left-Right models \cite{LR},
multiple Higgs\cite{NHDM}, supersymmetry
\cite{ana,susy,FeruglioEFT,ArHe}, Little Higgs models \cite{BurasLHvrai,LH}
and extra dimensions \cite{flat,GGST,warped}.
The expectations of an A4 flavour
symmetry, implemented in dimension six lepton flavour violating
operators (so a New Physics model
is not required), were explored in  \cite{FeruglioEFT}. 
 A more
model-independent approach \cite{ana,Paradisi},
uses a ``penguin-box'' expansion \cite{BurasHouches}, 
to classify  New Physics models  according to  
 the  dominant   effective
interactions which they  induce among ( three or four) SM particles.  
The  various  possibilities
predict correlations \cite{ana,ArHe,Paradisi,TarantinoLFV,susygut} 
among observables
(such as $(g-2)_\mu$ and $\meg$\cite{HT}, $\meee$ and $\meg$\cite{OOS},
or $\meg$ and $\mec$\cite{Kitano}).

This paper studies flavour structure in the  three
$\llg$ decays, which are induced by the three-particle
dipole vertex given in eqn (\ref{dipole}). Electric and magnetic
dipole moments are neglected because
they are flavour diagonal (see the expansion (\ref{Xmfv})),
and  other observables are usually  neglected
 because they are induced by
other operators.

Lepton flavour violating 
radiative
decays,  $e_\a \to e_\b \gamma$,
  proceed
via the operator 
(\ref{dipole})
at a rate given by 
\bea \label{radiative_decay}
\widetilde{BR} (e_{\alpha} \to e_{\beta} ~ \gamma)
\equiv \frac{\Gamma (e_{\alpha} \to e_{\beta} ~ \gamma)}
{\Gamma(e_\alpha \rightarrow e_\beta \nu_{\alpha} \bar{\nu}_{\beta})} 
&   =  & 
\frac{ \alpha_{em} m_\alpha^3}{256 \pi^4} (|X_{L  \b \a}|^2 + |X_{R  \b \a}|^2) \frac{192 \pi^3} 
{G_F^2 m_\alpha^5} ~~~, 
\eea
where $\a,\b \in \{ e, \mu,\tau \}$,
and  the  whole paper uses 
the charged lepton mass eigenstate
basis.
 Notice that
in table \ref{tab}, as in
the remainder of this paper, 
the $\tlg$ ``branching ratios with tilde'' are
defined  with respect to the leptonic
$\tau$ decay rate    to facilitate
comparaison with $\meg$. 
Table \ref{tab} lists the current bounds,  and hoped
for sensitivities of running or planned experiments.
  \begin{table}
\renewcommand{\arraystretch}{1.25}
$$\begin{array}{|c|c|c|}\hline 
\widetilde{BR} & \hbox{current ~ bound} & \hbox{future} \\
\hline
\meg & 2.4 \times 10^{-12}\cite{MEG} & \sim 10^{-13} , \hbox{(MEG \cite{MEG})} \\
\tmg &  2.5 \times 10^{-7} \cite{Babartmgteg,Belletmg} & \sim 10^{-8} , 
\hbox{(super-B~ factories\cite{SB})} \\
\teg &  1.9 \times 10^{-7}\cite{Babartmgteg} & \sim 10^{-8} , 
\hbox{(super-B~ factories\cite{SB})} \\
\hline
\end{array}$$
\caption{Current bounds  and hoped-for sensitivities
to lepton flavour violating branching ratios,  normalised
to leptonic weak decays, as 
in eqn (\ref{radiative_decay}). 
\label{tab}}
\renewcommand{\arraystretch}{1.00}
\end{table}

The  dipole vertex functions 
of  eqn (\ref{dipole}) satisfy
 ${\bf X}_R^\dagger = {\bf X}_L$:
$[{\bf X}_R]_{e \mu}$ induces
$\mu_R \to e_L \gamma$, 
$[{\bf X}_L]_{e \mu}$ induces
$\mu_L \to e_R \gamma$, 
$[{\bf X}_R]_{ \mu e}$ induces
$(\mu_L)^+ \to (e_R)^+ \gamma$, 
and $[{\bf X}_L]_{ \mu e}$  induces
$(\mu_R)^+ \to (e_L)^+ \gamma$.
Since we are interested in $\llg$ rates
(and not CP violation), 
we only need to consider    the $\a > \b$ 
components (lower triangle) of  $[{\bf X}_R]_{\b \a}$
and $ [{\bf X}_L]_{\b \a}$, which are independent. 
 The dipole vertex functions
can be  written as :
\bea
\label{Xmfv}
{\bf X}_R & =&   v \left( c_R  {\bf Y^e}
+  {\bf S}_{L}  {\bf Y^e}  
+   {\bf T}_R~ \right)  \frac{4 G_F }{\sqrt{2}} \\
{\bf X}_L  & = &  v  \left( c_L  {\bf Y^e}^\dagger
+  {\bf S}_{R}  {\bf Y^e}^\dagger
+    {\bf T}_L \right) \frac{4 G_F }{\sqrt{2}} 
~~,\nonumber
\eea
where $v = 174 $ GeV is the Higgs
vev.
 The normalisation by $2 \sqrt{2} G_F$ is concrete,
however,  the masses of the  new particles which mediate
$\llg$ will be assumed to  exceed the weak
scale $v$. 
 $ {\bf Y}_e$ is the (diagonal)  charged
lepton Yukawa matrix,
$c_L$ and $c_R$  are unknown, unflavoured
constants,
 and ${\bf S}_L$,
 ${\bf S}_R$  and  ${\bf T}_{L,R}$ are new  dimensionless matrices
in flavour space, which we will
refer to as ``flavour tensors''. 
Section \ref{sec:invar} 
 studies  when they can
be constructed by multiplying together 
flavoured matrices 
that one finds in  the Lagrangian. 
 ${\bf S}_L$ and 
 ${\bf S}_R$ are hermitian, 
respectively 
induced by New Physics which couples 
to doublet leptons,
or to singlet leptons. 
${\bf T}_R =\T_L^\dagger $ is an arbitrary matrix
(as was ${\bf X}$), ${\bf T}_R $ is defined
 like the Yukawas to have doublet-singlet
index order, induced by New Physics
interacting with doublet and singlet leptons.
The ${\bf X}_L, {\bf X}_R$
are labelled by the chirality of the incoming charged
lepton; the  $\SX$ have the opposite 
chiral label, because the incoming
chirality is flipped by a charged lepton 
Yukawa coupling. 

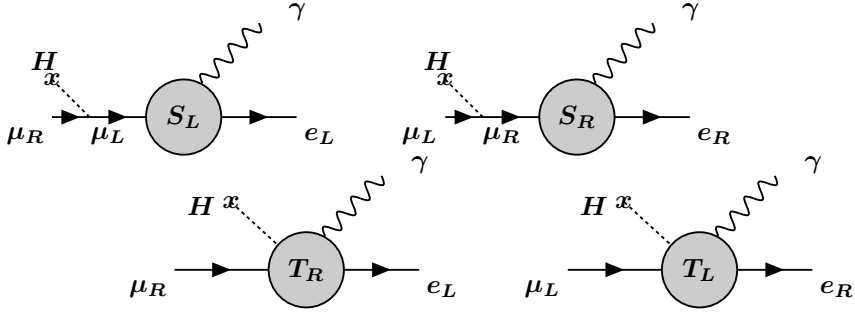
\begin{figure}[ht]
\unitlength.5mm
\SetScale{1.418}
\begin{boldmath}
\begin{center}
\begin{picture}(60,40)(0,0)
\ArrowLine(-5,0)(5,0)
\ArrowLine(5,0)(20,0)
\GCirc(30,0){10}{.8}
\DashLine(-5,10)(5,0){1}
\ArrowLine(40,0)(60,0)
\Photon(35,9)(50,25){2}{4}
\Text(-7,-5)[r]{$\mu_R$}
\Text(62,-5)[l]{$e_L$}
\Text(58,28)[l]{$\gamma$}
\Text(10,-5)[c]{$\mu_L$}
\Text(-7,15)[c]{$H$}
\Text(30,0)[c]{$S_L$}
\Text(-5,10)[c]{$x$}
\end{picture}
\quad \quad \quad \quad  \quad \quad
\begin{picture}(60,40)(0,0)
\ArrowLine(-5,0)(5,0)
\ArrowLine(5,0)(20,0)
\GCirc(30,0){10}{.8}
\DashLine(-5,10)(5,0){1}
\ArrowLine(40,0)(60,0)
\Photon(35,9)(50,25){2}{4}
\Text(-7,-5)[r]{$\mu_L$}
\Text(62,-5)[l]{$e_R$}
\Text(58,28)[l]{$\gamma$}
\Text(10,-5)[c]{$\mu_R$}
\Text(-7,15)[c]{$H$}
\Text(-5,10)[c]{$x$}
\Text(30,0)[c]{$S_R$}
\end{picture}
\newline
\begin{picture}(60,40)(0,0)
\ArrowLine(-5,0)(20,0)
\GCirc(30,0){10}{.8}
\DashLine(10,18)(22,7){1}
\ArrowLine(40,0)(60,0)
\Photon(35,9)(50,25){2}{4}
\Text(-7,-5)[r]{$\mu_R$}
\Text(62,-5)[l]{$e_L$}
\Text(58,28)[l]{$\gamma$}
\Text(2,17)[c]{$H$}
\Text(10,18)[c]{$x$}
\Text(30,0)[c]{$T_R$}
\end{picture}
\quad \quad \quad \quad  \quad \quad
\begin{picture}(60,40)(0,0)
\ArrowLine(-5,0)(20,0)
\DashLine(10,18)(22,7){1}
\GCirc(30,0){10}{.8}
\ArrowLine(40,0)(60,0)
\Photon(35,9)(50,25){2}{4}
\Text(-7,-5)[r]{$\mu_L$}
\Text(62,-5)[l]{$e_R$}
\Text(58,28)[l]{$\gamma$}
\Text(2,17)[c]{$H$}
\Text(10,18)[c]{$x$}
\Text(30,0)[c]{$T_L$}
\end{picture}
\end{center}
\end{boldmath}
\vspace{10mm}
\caption{The first blob corresponds to $[\SL]_{e \mu}$, 
the second to
$[\SR]_{e \mu}$. On the second row, the
blobs correspond to  $[\T_R]_{e \mu}$ and $[\T_L]_{e \mu}$.
\label{fig:ST}
}
\end{figure}
 The decomposition of
eqn (\ref{Xmfv})  corresponds to 
 flavour diagonal  + flavour changing new physics  
that is not chirality-flip (the $\SX$ terms  therefore  contains  $ Y_e$)
+  flavour changing chirality flip new physics represented
by $\T_{L,R}$. 
The first
 term $\propto {\bf Y}_e$, whose
coefficient can describe the  magnetic and electric dipole
moments,  can not induce the flavour off-diagonal 
$\llg$. 
This term is nonetheless included,
so that the diagonal
elements of our  new flavour
tensors are not required to
parametrise flavour diagonal new physics.
The lower triangles of 
$\SX$, $\T_L$ and $\T_R$ correspond
respectively to the four diagrams of figure
\ref{fig:ST}.  
We have  neglected  
the  possible $ {\bf Y}^\dagger_e \SL $ contribution to
${\bf X}_L$, and  the  $ {\bf Y}_e \SR $
contribution to ${\bf X}_R$, because
they are proportional to the outgoing charged lepton mass
$m_\b$.

In the presence of either the $\SX$ or $\T$ flavour
tensors \footnote{if both are present, there
are $ST$ interference terms}, the ratios of branching ratios
of eqn (\ref{radiative_decay})
become (recall that $4 G_F/\sqrt{2} = 1/v^2$, for
$v = 174$ GeV):
\bea
\widetilde{BR} (l_{\alpha} \to l_{\beta} ~ \gamma)
&   =  & 
\frac{ 6  \alpha}{ \pi}
  \left(    |S_{L  \b \a}|^2  + 
  |S_{R  \b \a}|^2\right)
\nonumber 
\\
& \sim&  1.4 \times  10^{-2} 
\left( 
 |S_{L  \b \a}|^2 +   
|S_{R  \b \a}|^2\right) 
\label{BRS}
\eea
\bea
\widetilde{BR} (l_{\alpha} \to l_{\beta} ~ \gamma)
&   \sim  &  
\frac{140  m_\tau^2}{  m_\alpha^2}
\left( 
 |T_{L  \b \a}|^2 +  
 |T_{R  \b \a}|^2\right) 
~~~.
\label{BRT}
\eea
If  the
the dimensionless couplings in  $T_{X  \b \a}$
and  $S_{X  \b \a}$ are taken  $\lsim 1$,  such that
 $S_{X  \b \a} \sim v^2/\Lambda_{NPFC}^2$ and  
$T_{X  \b \a} \sim v^2/\Lambda_{NPFC}^2$,  then 
Super-B factories could probe to 
a new flavour changing  scale  $\Lambda_{NPFC} \lsim 10$ TeV for $\SX$ and 
$\Lambda_{NPFC} \lsim 60$ TeV
for $\T$.

Convenient formulae for the $[X_X]_{\b  \a}$ vertex function,
generated in one-loop diagrams involving a boson and a 
fermion $f$ (see {\it e.g.} figures \ref{fig:W}  and \ref{fig:scal}),
are presented in the paper of Lavoura \cite{lavoura}.
 For a scalar  $S$ with Yukawa-type interactions  
\beq
{\cal L}_{ (Lavoura)} = S \overline{f}
(\lambda_{L \sigma} P_L + \lambda_{R \sigma}P_R) e_\sigma + h.c.
\label{Lscal}
\eeq
where $e_\sigma$ is  a  charged lepton of flavour $\sigma$, 
in four  component notation,
Lavoura obtains 
\bea
\label{Lav1}
 \frac{ [S_R]_{\b \a}}{v^2}& = & -\frac{  \rho_{\b \a} }{2m_S^2} \left[
Q_f k(t)  + Q_S  \overline{k}(t) \right]   \\
 \frac{[S_L]_{\b \a}}{v^2}& = &-\frac{  \omega_{\b \a}}{2m_S^2} \left[
Q_f k(t)  + Q_S  \overline{k}(t) \right]   \\
 \frac{[T_R]_{\b \a}}{v^2} & = &-\frac{ m_f \xi_{\b \a}}{2m_S^2v} \left[
Q_f k_f(t) + Q_S  \overline{k_f}(t) \right] \\
 \frac{[T_L]_{\b \a}}{v^2} & = &-\frac{  m_f \zeta_{\b \a} }{2m_S^2v} \left[
Q_f k_f(t) + Q_S  \overline{k_f}(t) \right]
\label{Lav4}
\eea
where the electric charges  satisfy
 $Q_S -Q_f = 1$,  the dimensionless functions $k, \overline{k}, k_f$  and
$ \overline{k_f}$ of
the mass ratio $ t = m_f^2/m_S^2$
 are given in the  Appendix,
and
\beq
\omega_{\b \a} = \lambda_{L \b}^* \lambda_{L \a}
~~~,~~~
\rho_{\b \a} = \lambda_{R \b}^* \lambda_{R \a}
~~~,~~~
\xi_{\b \a} = \lambda_{L \b}^* \lambda_{R \a}
~~~,~~~
\zeta_{\b \a} = \lambda_{R \b}^* \lambda_{L \a}
~~~.
\label{defnaa}
\eeq

Lavoura also gives formulae for a vector and fermion loop.
Suppose that  $W$ bosons  interact with $e_{L\sigma}$ and a
neutral fermion $f$  via the  Lagrangian
\beq
{\cal L}_{vec} = A'_{L \sigma} W_\rho \overline{f} \gamma^\rho e_{L\sigma}
+ h.c. + \delta {\cal L}_{F'H}
\label{Lvec}
\eeq
where $A_{L \sigma}^{'} = g/\sqrt{2} \times $mixing angles,
and    $\delta {\cal L}_{F'H}$ is the additional interactions,
in Feynman 'tHooft gauge, of the Higgs goldstone components.
Then Lavoura gives
\bea
\label{Lav5}
 \frac{ [S_R]_{\b \a}}{v^2}& = 
&- \frac{  \tilde{\rho}_{\b \a} }{2m_W^2} 
 Q_W  \overline{y}(t)   
~~~~,~~~~
{\rm where} ~~~\tilde{\rho}_{\b \a} = A_{L\b}^{'*}  A_{L\a}^{'}
 \eea
$\overline{y}$ is given  in  the  Appendix,
and  $[S_L]_{\b \a} = [T_R]_{\b \a} = [T_L]_{\b \a} = 0$.

In the $SU(2) \times U(1)$ invariant Lagrangian, 
there is a   dimension six  operator $O^{e \gamma}$,   
which contributes to the photon dipole operator,  and
which is a   linear combination
of    hypercharge and weak SU(2)  operators:
\beq
O^{e \gamma}_{\b \a} \equiv 
(\overline{\ell}_\b H) \sigma^{\mu \nu} P_R e_\a F_{\mu \nu}
=
\cos \theta_W (\overline{\ell}_\b H) \sigma^{\mu \nu} P_R e_\a B_{\mu \nu}
- \sin \theta_W
(\overline{\ell}_\b \tau^3  H) \sigma^{\mu \nu}  P_R e_\a {W}^3_{\mu \nu}
\label{v1}
\eeq
where $\ell$ is an SU(2) doublet, 
$\sigma ^{\mu \nu} = i [\gamma^\mu, \gamma^\nu]/2$,
and  $\{ \tau^i \}$ are the Pauli matrices.
If the operator of eqn (\ref{v1}) appears in the Lagrangian
normalised as  ${\cal L} \supset e 2 \sqrt{2} G_F  {\bf C}_{\b \a}^{(6)} 
O_{\b \a}^{e \gamma} + h.c.$, then
 ${\bf C}_{\b \a}^{(6)}$ contributes
to  $[{\bf X}_{R}]_ {\b \a}$ (see eqn (\ref{Xmfv})).
Notice that 
 there could also 
be higher dimensional dipole operators,
such as
\beq
O_{\b \a}^{e \gamma H^2} \equiv 
(H^\dagger H) (\overline{\ell}_\b H) \sigma^{\mu \nu} P_R e_\a F_{\mu \nu}
\eeq
 with  additional
Higgs legs and  suppressed by
more powers of the flavour changing  New Physics scale
$\Lambda_{NPFC}$.
Therefore,  ${\bf C}^{(6)}$   
corresponds to the ${\cal O} (1/\Lambda_{NPFC}^2)$ terms in   
${\bf X}_R$,  
and the coefficients of 
  $O^{e \gamma H^2}$   
include  ${\cal O} (m_{SM}^2/\Lambda_{NPFC}^4)$ 
contributions\footnote{The coefficient of 
 $O^{e \gamma H^2}$ would also include the contributions
$\propto v^2$ to the  mass$^2$ of the new particles.
However, if we propagate mass eigenstate new particles,
such contributions would remain resummed on the denominator.}
 (where $m_{SM}$ is a SM mass).   
In
this paper, we  assume that
there is a New Physics contribution
to  ${\bf C}^{(6)}$
and  we will be interested in
identifying cases where  ${\bf C}^{(6)} \simeq {\bf X}_R$.

\section{Invariants}
\label{sec:invar}

For the purposes of this paper, invariants
 are tensors in flavour space, often
a scalar or a matrix, obtained by multiplying flavoured matrices
from the Lagrangian  (``spurions'', in Minimal
Flavour Violation \cite{dAGIS} langauge).
They are interesting for Beyond the Standard Model physics,
because, if a sufficient number of invariants are measured, the
flavoured matrices of the Lagrangian can be
``reconstructed''  by simple matrix
multiplication.  Like the original
invariant of Jarlskog \cite{Jinvar}, these
invariants are also an elegant way to avoid dependance
on unphysical or unknown basis choices in the Lagrangian. So
the invariants in this paper are allowed to have flavour indices,
but these must be in the mass eigenstate
basis of known particles.

It is clear that  that the formulae (\ref{Lav1}) - (\ref{Lav4})
and  (\ref{Lav5})
have the makings of ``Jarlskog-like invariants'',
at least in the case where the $k$-functions can
be approximated by a single term.  For instance, if
$m_f\ll m_S$,
then $\SX \sim {\lambda} {\bf m}_S^{-2}{\lambda}^\dagger$.
This would be a coefficient of the dimension
six operator given in eqn (\ref{v1}).
The remaining terms from the $k$ functions, $\propto (m_f/m_S)^n$
for $n\geq 1$, would give coefficients for
dimension $>6$ operators which give more suppressed
 contributions to the
dipole vertex functions.  
The first aim of this paper, is to identify when
the flavour tensors  $\SX$, $\T_L$ and $\T_R$  are 
invariants.
Various questions come to mind about this
connection: what happens to logarithms? What
if both particles in the loop are flavoured?
What happens when there are several
new particles with slightly
different masses?
 Whether the 
vertex functions are invariants is essentially a ``top-down''
question.   However, it is phenomenologically
interesting, because we would like to reconstruct
the fundamental  Lagrangian from observables,
so it is useful to know when the connection
is simple.


\subsection{The non-supersymmetric Seesaw }
\label{sec:seesaw}

The  type I seesaw \cite{seesaw}, (without supersymmetry), 
 is simple model in which to study the invariant
question, because it only contributes to
$\SL$, and because  only one of the particles
exchanged in the loop (the neutral
fermion)  carries a generation
index.
The singlet neutrino masses are assumed
to be at the TeV-scale\cite{lowscale}, 
 as can for instance arise in  the
``inverse seesaw'' \cite{inverse}. 
Some recent  constraints on  such models
can be found in \cite{bounds,ABBGH}.
This mass scale is chosen so as 
to obtain detectable $\llg$ rates \cite{C+L1,ABCMP}, 
while  maintaining the new
mass scale above the Higgs vev $v$. 
(The seesaw behaviour
works for  singlet masses $\ll $ TeV
\cite{lightseesaw}, but our 
discussion will not apply.)

Section \ref{sec:seesawmod}  approximately
diagonalises the neutral mass matrix, and checks that 
the known  $\llg$ amplitudes are reproduced
\footnote{Despite  comments in \cite{ABBGH},
the correct
amplitudes can also be obtained by first matching
the unbroken gauge theory at the heavy scale $M$
onto the dimension six ``kinetic'' operator 
$\overline {\ell} H \dslash H^* \ell$ of \cite {GGST}
and  the dipole operator of eqn (\ref{v1}),  and  then
matching  at $m_W$, in the broken theory,
 onto the dipole, including the kinetic
operator in the loop. At the high-scale matching,
it seems that the not-1PI diagrams with a Higgs
insertion on the external leg should be included.}
using the formulae of \cite{lavoura} for
the neutral fermion and $W$  loop diagram.
Section \ref{sec:seesawinvar}  studies
the conditions under which the 
amplitudes are proportional to invariants.

\subsubsection{The model}
\label{sec:seesawmod}

Consider adding   $r$  heavy ($M_I \gsim$ TeV) 
singlet fermions $N$ , with a
majorana mass matrix  $M$,  to  the three generation SM.
In a simple inverse seesaw model,  there would be $r = 6$
singlets, arranged in three pairs of opposite-sign,
but almost-equal magnitude, masses. 
The leptonic Lagrangian, in  the mass basis of charged
leptons and heavy singlets, will be: 
\bea
\label{L}
{\cal L} & =&  \lambda_{\a I}\overline{\ell}_\a H^*_u  N_I
- \frac{M_{I}}{2} N_IN_I
-y_\a \overline{\ell}_\a H^*_d  e_\a + h.c. 
\eea
where $\overline{\ell} H_u^* = 
\overline{e} (H^+_u)^*
- \overline{\nu} (H_u^0)^*$, and two Higgs
doublets are allowed for.  The neutral mass terms  will be
\bea
{\cal L}_{mass} 
 & \to & -  \lambda_{\a I}  \overline{\nu_\a} 
\langle  H^{0 *}_u \rangle 
 N_I
- \frac{M_{I}}{2} N_IN_I
 + h.c. \nonumber 
\eea
and the  resulting  $(3+r) \times (3+r)$  majorana
mass matrix can be diagonalised with a unitary
matrix $X$:
\beq
X 
\left[
\begin{array}{cc}
0  & \lambda v_u \\
\lambda^T v_u & M
\end{array}
\right]
X^T = 
\left[
\begin{array}{cc}
D_m &0 \\
0 & D_M
\end{array}
\right]
~~~~{\rm where} ~~~
D_m \equiv -  v_u^2  U^\dagger \lambda M^{-1} \lambda^T U^*
\eeq
where $D_M = $ diag$\{ M_1,..M_{r}\}$,
and  we suppose that it is sufficient to obtain
$X$ to order $1/M^2$, because $O^{e\gamma}$ of
eqn (\ref{v1}) is a dimension six operator. 
Assuming that det [$M$] $\neq 0$, because all
the singlets are heavy,  and taking $U$ to be the
usual leptonic mixing matrix, gives
\beq
X = 
\left[
\begin{array}{cc}
U^\dagger  &0 \\
0 & 1
\end{array}
\right]
\left[
\begin{array}{cc}
1 - \delta_1& - \lambda M^{-1} \\
M^{-1 *} \lambda^\dagger & 
1 - \delta_2
\end{array}
\right]
\eeq
where the  $\delta_i$ ensure that $X$ is unitary
at $O(1/M^2)$:
\beq
\delta_1 = \frac{v_u^2}{2} \lambda [M^\dagger M]^{-1} \lambda^\dagger
~~~~, ~~~~
\delta_2 = \frac{v_u^2}{2}  [M^\dagger]^{-1}  \lambda^\dagger \lambda M^{-1}
\label{delta}
\eeq
A flavour eigenstate neutrino, participating
in a vertex with the $W$ and a charged lepton,
can therefore be expanded on light ($\nu_i$)  and
heavy ($n_J$) four-component majorana mass eigenstates as
\beq 
\label{nua}
\nu_\a = \sum_i {\Big (}
 U_{\a i} 
- \frac{v_u^2}{2} [\lambda [M^\dagger M]^{-1} \lambda^\dagger U]_{\a i}
{\Big )}
\nu_i
+  \sum_J \frac{ v_u \lambda_{\a J}}{M_J} n_J
\eeq

\subsubsection{To get invariants?}
\label{sec:seesawinvar}

\begin{figure}[ht]
\unitlength.5mm
\SetScale{1.418}
\begin{boldmath}
\begin{center}
\begin{picture}(60,40)(0,0)
\ArrowLine(-5,0)(5,0)
\ArrowLine(5,0)(45,0)
\ArrowLine(45,0)(60,0)
\Photon(15,0)(30,15){2}{3}
\Photon(45,0)(30,15){2}{3}
\Photon(30,15)(50,30){2}{5}
\Text(-7,-5)[r]{$\mu$}
\Text(62,-5)[l]{$e_L $}
\Text(17,13)[c]{$W$}
\Text(47,12)[c]{$W$}
\Text(30,-5)[c]{$n_I,\nu_i $}
\end{picture}
\qquad\qquad
\begin{picture}(60,40)(0,0)
\ArrowLine(-5,0)(15,0)
\ArrowLine(15,0)(45,0)
\ArrowLine(45,0)(60,0)
\DashCArc(30,0)(15,0,180){1}
\Photon(40.6,10.6)(60,30){2}{5}
\Text(-2,-5)[r]{$\mu$}
\Text(62,-5)[l]{$e_L$}
\Text(30,20)[c]{$H$}
\Text(30,-5)[c]{$n_I,\nu_i $}
\end{picture}
\qquad\qquad
\begin{picture}(60,40)(0,0)
\ArrowLine(-5,0)(15,0)
\ArrowLine(15,0)(45,0)
\ArrowLine(45,0)(60,0)
\DashLine(15,0)(30,15){1}
\Photon(45,0)(30,15){2}{3}
\Photon(30,15)(50,30){2}{5}
\Text(-7,-5)[r]{$\mu$}
\Text(62,-5)[l]{$e_L$}
\Text(20,15)[c]{$H$}
\Text(45,10)[c]{$W$}
\Text(30,-5)[c]{$n_I,\nu_i $}
\end{picture}
\qquad\qquad
\begin{picture}(60,40)(0,0)
\ArrowLine(-5,0)(15,0)
\ArrowLine(15,0)(45,0)
\ArrowLine(45,0)(60,0)
\Photon(15,0)(30,15){2}{3}
\DashLine(45,0)(30,15){1}
\Photon(30,15)(50,30){2}{5}
\Text(-7,-5)[r]{$\mu$}
\Text(62,-5)[l]{$e_L$}
\Text(20,15)[c]{$W$}
\Text(45,10)[c]{$H$}
\Text(30,-5)[c]{$n_I,\nu_i $}
\end{picture}
\end{center}
\end{boldmath}
\vspace{10mm}
\caption{One-loop diagrams that  contribute to  
the dipole vertex function 
in the seesaw model with broken electroweak symmetry. $H$ is the goldstone. 
A Higgs leg, which is not drawn, can attach to
the $\mu$ or $n_I$ line.
\label{fig:W}}
\end{figure}
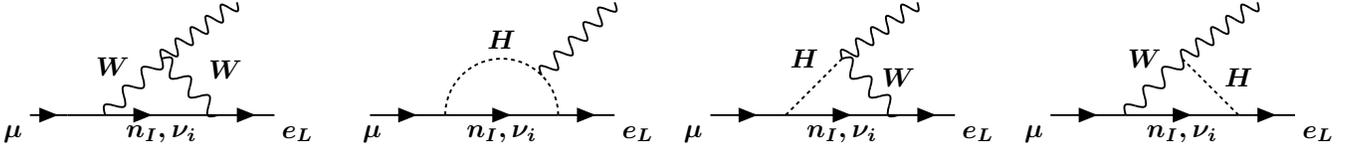

The contribution to the dipole flavour tensors
is obtained by summing the diagrams where a
$W$ boson  and a light or heavy neutral
fermion is exchanged. The relevant diagrams
are given in figure  \ref{fig:W}.
 We neglect the light neutrino
mass$^2$ contribution because it is parametrically
of ${\cal O}(\lambda^4)$, to be compared
to the ${\cal O}(\lambda^2)$ contributions
from heavy neutral exchange and from
the ``non-unitarity''  of the light neutrals.

Combining  eqn (\ref{Lav5}) with eqn (\ref{nua})
gives
\bea
\frac{[S_R]_{\b \a}}{v^2} & = & - \frac{g^2v_u^2}{4m_W^2}
[ \lambda [M^\dagger M]^{-1} \lambda^\dagger]_{ \b \a}
{\Big (}
-  \overline{y}_1(0) + 
 \overline{y}_1(M^2_J/m_W^2)
{\Big )}  \nonumber \\
  & = & -
\frac{ s_\b^2 }{4}
[ \lambda [M^\dagger M]^{-1} \lambda^\dagger]_{ \b \a}
~+...
~~~ ~~~ ~~~ {\rm heavy + light ~ in~ loop}
\label{SRseesaw}
\eea
where  $v_u/v \equiv s_\b$ and $\overline{y}$
is given in eqn (\ref{y1bar}), and only
the  ${\cal O}(1/M^2)$ terms were retained.
This is consistent with calculating the diagonalisation
matrix $X$ to  ${\cal O}(1/M^2)$.

As hoped for, 
equation (\ref{SRseesaw}) is an invariant constructed from
flavoured matrices from the Lagrangian. This is
the case despite electroweak symmetry breaking
and the separate contributions from the
 scales $M$ and $m_W$. 
 However, there are   ${\cal O}(v^2/M^4)$
contributions from higher order terms in
the expansion of $\overline{y}$. From eqn
 (\ref{y1bar}),   one sees that heavy and light
neutral exchange will give contributions
with coefficients proportional to
\beq
v^2 [ \lambda [M^\dagger M M^\dagger M]^{-1} \lambda^\dagger]
~~~... ~,~~~
[ \lambda M^{-1} \lambda^T \lambda^*  M^{-1 *} \lambda^\dagger]
~,~
v^2 [ \lambda [M^\dagger  M]^{-1} \lambda^\dagger
 \lambda M^{-1} \lambda^T \lambda^*  M^{-1 *} \lambda^\dagger]
~~~...
\label{autres}
\eeq
Since the eigenvalues and eigenvectors
of the neutral mass matrix   were only obtained 
to ${\cal O}(1/M^2)$, 
 the  numerical factors
multiplying the   ${\cal O}(v^2/M^4)$ terms
are unknown.  The  coefficients are still invariants,
which seems reasonable when the   flavoured 
 particles in the loop 
 all have masses beyond current experimental reach. 
Fortunately, there are no logarithms at dimension six.

\subsection{Scalar leptoquarks}
\label{sec:LQ}

The second toy model is various
scalar leptoquarks. For updated low energy
 bounds, and
references to earlier works, see {\it e.g.}  \cite{LQrev}.  
The current  collider  lower bound on   the mass
of a leptoquark decaying to first generation leptons
and quarks is 660 GeV \cite{ATLASslides},
based on 1 $fb^{-1}$ of data. See \cite{LQolder}
for earlier collider mass bounds.

The scalar leptoquark contribution
to the dipole vertex function is more
complicated 
than the seesaw model, because they
 can contribute to $\SL, \SR$, or $\T$
(because they can interact with both
singlet and doublet charged leptons),
and in that  both they and the quark
in the loop can be flavoured.

Subsection \ref{sec:LQL} gives 
the Lagrangian for baryon and lepton
number conserving  scalar leptoquarks, and the contributions
to $\llg$ of  two 
 SU(2) singlet
leptoquarks upon which we focus. 
 These are representative
of the possibilities (see table \ref{tab:LQ}),
because the $S_0$ leptoquark can interact
with singlet and doublet fermions.
Then subsection 
 \ref{sec:LQinvar}
discusses the  prospects for
obtaining invariants.

\subsubsection{The models}
\label{sec:LQL}

 Consider 
 scalar leptoquarks, 
with renormalisable $B$ and $L$
conserving interactions,
which can be SU(2) singlets, doublets
or triplets.  
In the notation of
Buchmuller,R\"uckl and Wyler\cite{BRW},
these
can be added to the SM Lagrangian  as:
\bea
{\cal L}_{LQ} & = & 
S_0 ( {\bf \lambda_{L S_0}} \overline{\ell} i \tau_2 q^c
+ {\bf \lambda_{R S_0}} \overline{e} u^c ) +
\tilde{S}_0 {\bf \tilde{\lambda}_{R \tilde{S}_0}} 
\overline{e} d^c   \nonumber \\ &&
+
 ( {\bf \lambda_{L S_2}} \overline{\ell}  u
+ {\bf \lambda_{R S_2}}  
\overline{e } q [i \tau_2  ])S_{2} +
 {\bf \tilde{\lambda}_{L \tilde{S}_2}} 
\overline{\ell} d \tilde{S}_2
 \nonumber \\ &&
+
 ( {\bf \lambda_{L S_3}} \overline{\ell} i \tau_2 \vec{\tau} q^c) 
\vec{S_3}
+ h.c.
\label{BRW}
\eea
where the $\lambda$s are 3 $\times$ 3 matrices 
with index order lepton-quark, and  $\{ \tau_i \}$ are Pauli
matrices, so $i \tau_2$ provides the antisymmetric SU(2) contraction.
In this Lagrangian, the leptoquark leaves the vertex into
which enter  the leptons.  This is converse to  Lavoura
conventions, where scalar and lepton both enter, 
and the internal fermion
leaves. 
Comparing eqn (\ref{BRW}) with  eqn (\ref{Lscal})
gives the parameters listed in table \ref{tab:LQ}.
  \begin{table}
\renewcommand{\arraystretch}{1.25}
$$
\begin{array}{l|cccccc}
{\rm leptoquark}  & Q_{S} \,   &  Q_f ~ (f)  & \omega  & \rho & \xi & \zeta  \\ 
\hline
\tilde{S}^\dagger_0 & 4/3       &1/3 ~ (d^c) 
& 0& 
[ \tilde{\lambda}_{R \tilde{S}_0}]_{\b f}
  [ \tilde{\lambda}_{R \tilde{S}_0}]^*_{\a f} 
  & 0 &0  \\
S^\dagger_0 & 1/3 &  -2/3~  (u^c) & 
 [ \lambda_{L S_0}]_{ \b q} [ \lambda_{L S_0}]^*_{ \a q}  & 
 [ \lambda_{R S_0}]_{ \b f} [ \lambda_{R S_0}]^*_{ \a f}&
 [ \lambda_{L S_0}]_{ \b f} [ \lambda_{R S_0}]^*_{ \a f}  &  
 [ \lambda_{R S_0}]_{ \b f} [ \lambda_{L S_0}]^*_{\a f}\\
S_{2}^\dagger (lower) & 5/3  &  2/3 (u) &  
[ \lambda_{L S_2}]_{\b f} [ \lambda_{L S_2}]^*_{\a f}   &
[ \lambda_{R S_2}]_{\b f} [ \lambda_{R S_2}]^*_{\a f} &
-[ \lambda_{L S_2}]_{\b f} [ \lambda_{R S_2}]^*_{\a f} &
-[ \lambda_{R S_2}]_{\b f} [ \lambda_{L S_2}]^*_{\a f} \\
S_{2}^\dagger (upper)  & 2/3 &  -1/3 (d) & 0  &
[ \lambda_{R S_2}]_{\b f} [ \lambda_{R S_2}]^*_{\a f} &
0  & 0 \\ 
\tilde{S}_2^\dagger (lower) &2/3   & -1/3 (d_R)  & 
 [ \tilde{\lambda}_{L \tilde{S}_2}]_{\b f} [ \tilde{\lambda}_{L \tilde{S}_2}]^*_{\a f}
 &  0  & 0
&0\\
\vec{S_3}^\dagger (\tau_3 comp) & 1/3 &  -2/3~  (u^c) & 
 [ \lambda_{L S_3}]_{ \b f} [ \lambda_{L S_3}]^*_{ \a f}  &
0&0&0\\
\end{array} 
$$
\caption{ Parameters for obtaining the $\llg$ amplitude,
induced by the scalar leptoquarks of the
left colomn, using eqns (\ref{Lav1}) - (\ref{Lav4})). 
$\alpha$ and $\beta$ are lepton flavour indices, and $f$ is
the internal quark index.  See figure \ref{fig:scal}. 
\label{tab:LQ}}
\renewcommand{\arraystretch}{1.00}
\end{table}
We will also allow the possibility  of three
generations \cite{LQgen} of  leptoquarks,
in whiich case
 $I:1..3$, and
$\lambda_{l q}^I$ is a three-index tensor.

\begin{figure}[ht]
\unitlength.5mm
\SetScale{1.418}
\begin{boldmath}
\begin{center}
\begin{picture}(60,40)(0,0)
\ArrowLine(-5,0)(5,0)
\ArrowLine(5,0)(15,0)
\DashLine(-5,10)(5,0){1}
\ArrowLine(15,0)(45,0)
\ArrowLine(45,0)(60,0)
\DashCArc(30,0)(15,0,180){2}
\Photon(45,15)(60,30){2}{4}
\Text(-7,-5)[r]{$e_\a$}
\Text(62,-5)[l]{$e_\b$}
\Text(62,30)[l]{$\gamma$}
\Text(30,20)[c]{$S$}
\Text(30,-5)[c]{$f$}
\Text(-7,15)[c]{$H$}
\Text(-5,10)[c]{$x$}
\end{picture}
\quad \quad \quad \quad
\begin{picture}(60,40)(0,0)
\ArrowLine(0,0)(15,0)
\DashLine(5,-10)(20,0){1}
\ArrowLine(15,0)(45,0)
\ArrowLine(45,0)(60,0)
\DashCArc(30,0)(15,0,180){2}
\Photon(45,15)(60,30){2}{4}
\Text(-2,-5)[r]{$e_\a$}
\Text(62,-5)[l]{$e_\b$}
\Text(62,30)[l]{$\gamma$}
\Text(30,20)[c]{$S$}
\Text(30,-5)[c]{$f$}
\Text(8,-15)[c]{$H$}
\Text(5,-10)[c]{$x$}
\end{picture}
\end{center}
\end{boldmath}
\vspace{10mm}
\caption{One loop  diagrams  contributing to $\llg$,
mediated by a scalar   $S$,  and a
fermion $f$. In the diagrams on
the left, the scalar $S$ interacts either
with doublet or singlet charged leptons,
so the Higgs insertion is
on an external leg.  In the diagrams
on the right, $S$ interacts with
singlet and doublet charged leptons,
so the Higgs insertion can be
on the internal lines. 
\label{fig:scal}}
\end{figure}
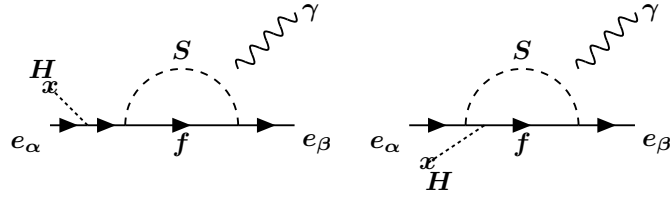

The SU(2) {\bf singlet leptoquark 
 $\widetilde{S}_0$},  which  couples to $\overline{d^c} e$,
contributes
to $\SR$ via the two diagrams summarised
by the left diagram  in  figure \ref{fig:scal} (the photon
can attach to $S$ or $f$),
where the internal fermion $f$ can be a singlet  $d^c,s^c$ or $b^c$
quark. For a $\widetilde{S_0}$ leptoquark,
eqn (\ref{Lav1}) implies 
that $\SL = \T_L = \T_R = 0$, and
\bea
\frac{[\SR]_{\b \a}}{v^2}  &= & 
- \sum_f\left[ \frac{ \rho_{\b \a}}{2 M_S^2} 
\left( k(x) + 4 \overline{k}(x)\right) \right]
\simeq  - \frac{1 }{4 M_S^2}
[ \tilde{\lambda}_{R \tilde{S}_0} 
   \tilde{\lambda}^\dagger_{R \tilde{S}_0}]_{\b \a } 
~~~, ~~~{\rm leptoquark~without~generations}
\label{Stinvar}
\eea
where  $x = m_f^2/M_{\tilde{S}}^2 \ll 1$.
Inside the square brackets after the first equality, 
$\rho_{\b \a}$  is from table \ref{tab:LQ} 
but not summed over quark flavour $q$.
Assuming  $x$ is negligeable, $k \to 1/6$
and $\bar{k} \to 1/12$, which gives the
second equivalence;
this means that the coefficients of dimension eight
operators, in this model, are irrelevant.

In the case where $\widetilde{S_0}$ has generation indices
$I$, one obtains
\bea
\frac{[\SR]_{\b \a}}{v^2}  &\simeq & - 
\frac{1 }{4}
\sum_I 
[ \tilde{\lambda}^I_{R \tilde{S}_0} 
\frac{1}{M_{S,I}^{2}}
   \tilde{\lambda}^{I \dagger}_{R \tilde{S}_0}]_{\b \a } 
\label{Stinvargen}
\equiv
- \frac{1 }{4}[ \tilde{\lambda}_{R \tilde{S}_0} 
 D_M^{-2}
   \tilde{\lambda}^\dagger_{R \tilde{S}_0}]_{\b \a } 
\eea
where after the first equality,
 there is a
$[\lambda^I]$  matrix with indices
in lepton and quark flavour spaces for each flavour
$I$ of leptoquark,
and  after the second equivalence, the leptoquark
index is also implicit 
 ($D_M^2 = $ diag $\{ M^2_{\tilde{S},1}..M^2_{\tilde{S},3} \}$
is the diagonal leptoquark
mass matrix).  $\lambda_{R \tilde{S}_0}$ is in the charged lepton, down quark
and leptoquark mass eigenstate bases, and
 the quark flavour sum has again been done
neglecting quark mass effects $\propto  m_q^2/M_{\tilde{S},I}^2$.

Consider now the
{\bf singlet ${S}_0$ coupling to $\overline{u^c_{L,R}} e_{L.R}$}, 
  without any flavour index of
its own, because it is
straightforward to add a flavour sum
on the leptoquarks to the formulae
below.   $S_0$ couples to up-type
quarks, so $ f = u,c,t$, and  the ${\cal O}(m_u^2, m_c^2)/M_S^2$
terms in $k$ and $\bar{k}$ can be neglected, as were the
quark mass terms  in the case of $\widetilde{S_0}$.
We keep  the top quark mass dependance.
For $x =m_f^2/M_S^2$, $t = m_t^2/M_S^2$, 
one obtains \cite{lavoura} 
\bea
\frac{[\SR]_{\b \a}}{v^2}  &= &  
-\sum_f \frac{ \rho_{\b \a}}{2M_S^2} \left( -2k(x) +  \overline{k}(x)\right)
 \nonumber \\ 
&\simeq & 
 \frac{ {\Big [} [ \lambda_{R S_0}] 
[ \lambda_{R S_0}]^\dagger {\Big ]}_{ \b \a}}{8M_S^2} 
- \frac{[ \lambda_{R S_0}]_{ \b t} [ \lambda_{R S_0}]^*_{ \a t}}{2M_S^2}
\left(   \frac{t^3 - 3t^2 + 9t}{4(t-1)^3} 
- \frac{t(t+2) \ln t}{2 (t-1)^4} \right)
\label{SRS0}
\eea
\bea
\frac{[\SL]_{\b \a}}{v^2}
 &= & 
-\sum_f \frac{  \omega_{ \b \a} }{2M_S^2} \left( -2k(x) +  \overline{k}(x)\right)
 \nonumber \\
&\simeq & 
 \frac{  {\Big [} [ \lambda_{L S_0}] [ \lambda_{L S_0}]^\dagger {\Big ]}_{ \b \a}}{8M_S^2} 
- \frac{[ \lambda_{L S_0}]_{ \b t} [ \lambda_{L S_0}]^*_{ \a t}}{2M_S^2}
\left(   \frac{t^3 - 3t^2 + 9t}{4(t-1)^3} - \frac{t(t+2) \ln t}{2 (t-1)^4} 
\right)
\label{SLS0}\\
\frac{[\T_R]_{\b \a}}{v^2}& = & 
- \sum_f
 \frac{ \xi_{\b \a}  m_f}{2 M_S^2 v} \left( -2k_f(x) +  \overline{k_f}(x)\right)
\nonumber  \\
&\simeq & -  \frac{ [ \lambda_{L S_0}]_{ \b f} m_f [ \lambda_{R S_0}]^*_{ \a f}}
{2 M_S^2 v} \left(   \frac{7}{2} + 2 \ln \frac{m_f^2}{M^2_S} \right) 
\nonumber  \\
& & ~~~ 
 -  \frac{ [ \lambda_{L S_0}]_{ \b t} m_t [ \lambda_{R S_0}]^*_{ \a t}}
{2M_S^2 v}
\left(   \frac{7t^2 + 13 t}{2(t-1)^2} 
- \frac{(2t^3 - 6t^2 + 7t) \ln t}{ (t-1)^3} 
\right)
 \label{TRS0} \\
\frac{[\T_L]_{\b \a}}{v^2}& = & 
-\sum_f
 \frac{\zeta_{\b \a} m_f}{2 M_S^2 v} \left( -2k_f(x) +  \overline{k_f}(x)\right)
\nonumber  \\
&\simeq & 
 -  \frac{[ \lambda_{R S_0}]_{ \b f} m_f [ \lambda_{L S_0}]^*_{ \a f}}
{2M_S^2 v}
 \left(   \frac{7}{2} +  2 \ln \frac{m_f^2}{M^2_S} \right) 
\nonumber  \\
& & ~~~  
-  \frac{ [ \lambda_{R S_0}]_{ \b t} m_t [ \lambda_{L S_0}]^*_{ \a t}}
{2M_S^2 v}
\left(   \frac{7t^2 + 13 t}{2(t-1)^2} 
- \frac{(2t^3 - 6t^2 + 7t) \ln t}{ (t-1)^3} 
\right) 
 \label{TLS0} 
\eea
where $\rho,\omega, \xi$ and $\zeta$ are from table
\ref{tab:LQ} with no sum over $q$.

\subsubsection{Getting invariants}
 \label{sec:LQinvar}

 For  the  $\widetilde{S_0}$, which interacts
with charged leptons and down-type quarks, 
   the dipole vertex function
is  given  by eqn  (\ref{Stinvar}), or
(\ref{Stinvargen}), respectively
for the cases that 
$\widetilde{S_0}$ does not, or does, have  generation
indices. 
It is an  invariant,  constructed from
spurions from the Lagrangian: 
\beq
\frac{[\SR]_{\b \a}}{v^2} = 
-
\frac{1}{4}
[ \tilde{\lambda}_{R \tilde{S}_0}
D_M^{-2} 
\tilde{\lambda}^{ \dagger}_{R \tilde{S}_0}]_{\b \a }
~~~, ~~~{\rm leptoquark}~
\widetilde{S_0}
{\rm ~with~generations}
\label{SRLAMSt0}
\eeq
where the quark flavour sum is implicit. 
 Since the down-type
quark masses are small compared to $M_{\widetilde{S}}^2$,
dimension eight and higher operators
are irrelevant,  the dipole
vertex function in this case is
unambiguously an invariant.

The dipole vertex functions generated
by the  leptoquark $S_0$, which couples to up-type
quarks,   are not so
obviously invariants due to the large
top mass.
For $\SX$, if 
 $9 \lambda_{\b t} \lambda^{*}_{\a t}  m_t^2/M_{S}^2 \ll 
[\lambda \lambda^{ \dagger}]_{\b \a }$,
  the second
term in eqns  (\ref{SRS0}) and (\ref{SLS0})  can be
 ignored,  and the ${\bf S}_X $ are approximately 
the first term, which is an invariant:
\beq
\frac{{\bf S}_X}{v^2} =  \frac{
{\Big [ } [ \lambda_{X S_0}] [ \lambda_{X S_0}]^\dagger 
{\Big ] }_{ \b \a}}{8M_S^2}
+...
\label{quisuisje}
\eeq
However, if the  terms of order  $ m^2_t/M_S^2$ 
should be included, they correspond to
dimension eight operators with invariant coefficient
$\propto \lambda Y_u Y_u^\dagger \lambda^\dagger$
(in the one-leptoquark-generation case).
This ressembles the   seesaw case, where
 the coefficients
of dimension eight operators
were  potentially dangerous, because
not neccessarily suppressed
by small couplings.

 In the presence of $\lambda_L$ and 
$\lambda_R$,  a contribution to $\T$ can arise,
propertional to the internal fermion mass,
giving the lowest order coefficient proportional
to invariants of the form $\lambda_R Y^\dagger _u \lambda^\dagger_L$ or
 $\lambda_L Y_u \lambda^\dagger_R$.
 So one obtains
\bea
\frac{[\T_L]_{\b \a}}{v^2}&  =& 
- 
\left(
\frac{1}{  M_{S}^2}
[ {\lambda}_{R {S}_0} \, ( c_1  Y_u^\dagger
+  Y_u^\dagger \ln ( Y_u^\dagger Y_u))
{\lambda}^{ \dagger}_{L \tilde{S}_0}]_{\b \a }
\right)  ~ + ...~~, \nonumber \\
\frac{[\T_R]_{\b \a}}{v^2}   & = &
- 
 \left(
\frac{1}{  M_{S}^2}
[ {\lambda}_{L {S}_0} 
( c_1  Y_u
+ Y_u \ln ( Y_u Y_u^\dagger ))
{\lambda}^{ \dagger}_{R\tilde{S}_0}]_{\b \a }
\right)  ~ + ...~~
~~~, ~~~{S_0}  ~{\rm with} ~\lambda_L , \lambda_R
\label{TLAMS0}
\eea
where $c_1 = 7/4 + \ln (v^2/M^2_S)$,  and  
the quark flavour sum is implicit. 
The logarithm is unfortunate, from
the perspective of reconstructing
the new  flavour structures, because
one needs its  coefficient 
to reconstruct the new flavoured
matrices. For instance, imagine
that $[\lambda_L]_{\b q}$
was determined in some other process,
and that  the pattern of $\llg$ decays
indicates that they are mediated
by a hierachical flavour tensor with
chirality flip on the internal line. 
Then  to obtain $\lambda_R$ from $\llg$ data
requires knowing the coefficient
of the  logarithm. Two positive
feature of logarithms are that one can
often guess their  presence, 
and, as  discussed in 
section 3.2.3.1 of \cite{ybook}, if one
knows the coefficient, 
the logarithm does not obstrct the reconstruction
of new flavour structures.

Eqn (\ref{TLAMS0}), like eqn (\ref{quisuisje}),
 neglects the dimension
eight  terms  of order
$ \lambda_R Y_u Y_u^\dagger  Y_u \lambda_L v^2/M_S^4$
and
$ \lambda_L Y_u Y_u^\dagger  Y_u \lambda_R v^2/M_S^4$,
which are not small for the top Yukawa.

Finally, it is clear that the dipole coefficient
induced by $S_0$ will only  be proportional
to a single invariant, if one 
of $\SX$, $\T_L$ or $\T_R$ dominates over the others.
See the comparaison of their
relative sizes  in section \ref{sec:LQhier}.

\section{Testing hierarchies?}
 \label{sec:hier}

It is well-known, that
if the flavour tensors $\SX$ or $\T$ are dominated by
their largest eigenvalue, then they are
compartively predictive. 
The idea is that, if
one of the  $\tlg$  decays is observed,
then  the current bound on $\meg$ implies 
 a small mixing angle in 
the diagonalisation of the flavour tensor,
and this small angle
 suppresses the other $\tau \to \ell'
\gamma$ decay.  This section is
somewhat independent of the previous
section, because the flavour tensors
contributing to the  vertex function can be
hierarchical without being invariants. Nonetheless,
subsection \ref{sec:hieretinvar} discusses when hierarchical
invariants are obtained, because ideally,
one wants to reconstruct the  spurions in the Lagrangian
from measurements of the vertex function.


\subsection{Defining hierarchies}
 \label{sec:notnhier}

We define a matrix, 
in particular the flavour tensors $\SX$, $\T_L$ and
$\T_R$,  to be ``hierarchical'', when the
off-diagonal elements are dominated
by the largest eigenvalue.  
A hermitian matrix 
 ${\bf S}$  (for instance $\SR$ or $\SL$),
can be  written
\beq
{\bf S} = V^\dagger D_S V ~~~, 
~~~~D_S = {\rm diag} \{ s_1, s_2, s_3 \} ~~~,~~~ s_1 \leq s_2 \leq s_3  
\label{Sdiag}
\eeq
and  $s_3$ will give the largest contribution
to  $S_{\b \a}$, when
\beq
 V_{1 \b}^* s_1V_{1 \a}~ ,~  V_{2 \b}^*  s_2 V_{2 \a}\ll
V_{3 \b}^*  s_3  V_{3 \a}
\label{cdn1}
\eeq
Eqn (\ref{cdn1})
is satisfied for eigenvalues $s_1,s_2$ and  $s_3$,
if the  mixing angles are large enough
\footnote{ Equation
(\ref{suffgrand}) implies
 (\ref{cdn1}), but since eqn (\ref{cdn1})
is a lower bound on a product of elements
of $V$, it can be satisfied when one
of the $V_{i j}$ does not satisfy eqn 
(\ref{suffgrand}).}: 
\beq
V_{j \b} \gsim \sqrt{ \frac{s_i}{s_j}} V_{i \b}  ~~~~i  \leq j ~~~. 
\label{suffgrand}
\eeq
This relates elements of a colomn in $V$:
the bottom row $[V]_{3 \a}$ has to be
bigger than a fraction $
\sqrt{s_i}$
 of the elements
in the colomn above it.

The flavour tensor ${\bf T}_R =  {\bf T}_L^\dagger$, like a  Yukawa matrix,
can be  diagonalised
with independent unitary matrices on
the left and right:
\beq
{\bf T}_{R} = V_L^\dagger D_T V_R ~~~, 
~~D_T = {\rm diag} \{ t_1, t_2, t_3 \} ~~~,~~~ t_1 \leq t_2 \leq t_3 
\label{Tdiag}
\eeq
If the matrix elements of $V_L$ and
$V_R$ satisfy eqn
(\ref{suffgrand}) (replacing  $s_i \to t_i$), then,
just as in the case of $\SX$,  the
off-diagonal elements  $[\T_L]_{\b \a}$, and
$[\T_R]_{\b \a}$ will be dominated by
$t_3$.

\subsection{Observing a  hierarchy}
\label{sec:hiertest}

Suppose initially that New Physics induces 
 only one of
$\SL$ or $\SR$  (the
possible presence  of $\SL$ and $\SR$,
or $\T_{L,R}$
will be discussed  later). Then  
we  start with  two   assumptions \\
1) $~$ the flavour tensor ${\bf S}_X$
is hierarchical \\
2) $~$ one of the $\tlg$ decays  is
observed at Super-B factories. 
This is
neccessary for them to be able
to test anything. 
To be concrete, suppose that
 $\tmg$ is observed. 
 \\
Combined with table \ref{tab}, these assumptions fix
\beq
3 \times 10^{-3}
\lsim  |V_{3\tau}^* s_3 V_{3 \mu}| 
\lsim
10^{-3}
\label{voir}
\eeq
where $s_3 \sim \lambda^2v^2/M_{BSM}^2  \leq v^2/M_{BSM}^2$
(assuming that
$\SX$ is induced at dimension six by
New Particles with couplings $ \lambda \leq 1$). 
In addition, the non-observation of $\meg$ 
puts an upper bound 
\beq
\label{BRbound}
\frac{\widetilde{BR}(\meg)}{\widetilde{BR}(\tmg)}
=\frac{ |{ S}_{ e \mu }|^2}{|{ S}_{ \mu \tau }|^2}
=\frac{|V_{3 e}|^2}{|V_{3 \tau}|^2}
\lsim  10^{-4}
~~~. 
\eeq
With this bound on $V_{3e}$, the only
way that $\teg$ could
 detectable rate at Super-B factories,  
is if $V_{3 \mu} \simeq  V_{3e}$ are 
both small. However, from eqn
(\ref{voir}), this would require
$s_3 \geq .1$, or
$M_{BSM}\lsim 500$ GeV, which
is a low scale for 
new particles with ${\cal O}(1)$
couplings to charged leptons\footnote{
The 7 TeV LHC, with the current 
 5 $fb^{-1}$ of data,
 is surely sensitive to
leptoquarks of mass $\lsim 600$ GeV ---even
if they decay to $ \tau t$ \cite{DV} ---
and possibly also to heavy singlet fermions}.
If we make the further assumption\\
$$
3) ~{\rm  that}~  s_3 \sim \frac{\lambda ^2 v^2}{M_{BSM}^2}
\ll .1 ~~, {\rm that~is}, M_{BSM} \gsim 500 ~{\rm GeV}
\hspace{10cm}
$$
then we can conclude that Super-B factories
can falsify the assumption of  hierarchical
couplings in $\SL$ or $\SR$, by
detecting both $\tmg$ and $\teg$. 

\begin{figure}[ht]
\unitlength.5mm
\begin{center}
\epsfig{file=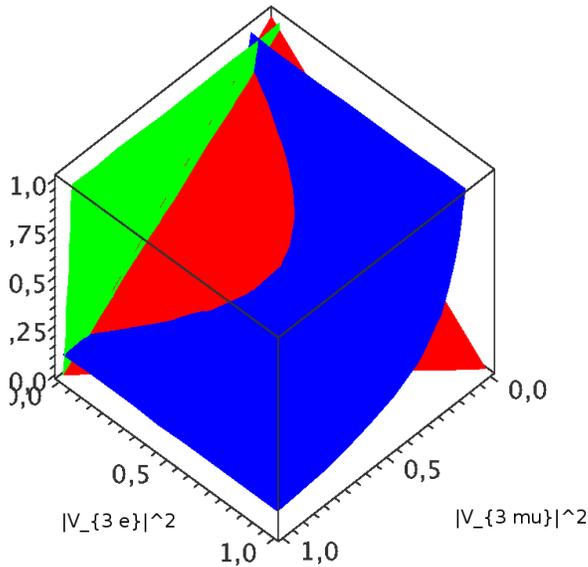,height=8cm,width=8cm}
\end{center}
\caption{ A graphical representation
of the  claim that a hierarchical
flavour tensor (see section \ref{sec:notnhier})
 predicts that a Super-B factory should
not see $\teg$ and $\tmg$.
The axes are 
 $\{|V_{3 e}|^2,|V_{3 \mu}|^2,|V_{3 \tau}|^2 \}$,
which satisfy $|V_{3 e}|^2 + |V_{3 \mu}|^2 + |V_{3 \tau}|^2 = 1$
represented by the red triangle. 
The vertical axis is $|V_{3 \tau}|^2$.
The blue hyperbola corresponds to 
$\Gamma(\tmg) \propto |V^*_{3 \mu} V_{3 \tau}|^2$
visible at a Super-B Factory. The
green plane corresponds to the $\meg$  bound
of eqn (\ref{BRbound}). Allowed points
live on the intersection of the
red and blue surfaces, and on the far
side of the green plane.
A hyperbola (not drawn)  corresponding
to $\teg$  at a rate similar
to $\tmg$ would live on the near side of
the green plane, see discussion after
assumption 3), so a Super-B Factory
should not see $\teg$ and $\tmg$. 
\label{fig:trimeg} }
\end{figure}

An alternative way to see the $\meg$ bound, for a hierarchical
flavour tensor,  is illustrated
in figure \ref{fig:trimeg}. 
The paremeter space of  the three elements
$|V_{3 \a}|^2$, who satisfy $\sum_\a
|V_{3 \a}|^2 = 1$, is  the red equilateral triangle.
The blue hyperbola corresponds to 
$\Gamma(\tmg) \propto |V^*_{3 \mu} V_{3 \tau}|^2$
visible at a Super-B Factory. Then the
green plane corresponds to the bound
of eqn (\ref{BRbound});  allowed points
live on the intersection of the
red and blue surfaces, and on the far
side of the green plane.
The hyperbola   corresponding
to  a fixed rate for $\teg$ is not
drawn  (it would be in the vertical
and $|V_{3e}|^2$ plane, intersecting
the blue hyperbola).  
 The green
and red planes  are independent of
the  (unknown) eigenvalue $s_3$,
which is $\propto v^2/M_{BSM}^2$. 
The hyperbolae do depend on $s_3$,
and
move away from the axes as $s_3$
decreases (or as 
the branching ratio increases for fixed
$s_3$).  For the value
of $s_3$ chosen in this plot, it is
clear that the $\teg$ hyperbola corresponding
to  a rate similar
to $\tmg$, would live on the near side of
the green plane, so a Super-B Factory
cannot see $\teg$ and $\tmg$.



The  assumption that only
one of $\SL$ or $\SR$ is present is minimal,
and perhaps justified by the  absence
to date of  new TeV-scale particles. 
However, it can in principle be tested:
if the decaying lepton is polarised,
then by angular momentum conservation,
the angular distribution of the final
lepton will depend on its chirality \cite{KunoOkada}. 
Therefore, if $\tmg$ and $\teg$ are both observed
at Super-B factories (where polarised
$\tau$s may be possible\cite{SBatVal,CDRSB}), 
one can hope to determine, from the
 final
state angular distributions, whether they are
induced by ${\bf S}_L$ and/or  ${\bf S}_R$.
If the final state angular distributions
are different, then this is compatible with 
 hierarchical   ${\bf S}_L$ and  ${\bf S}_R$;
if both $\tmg$ and $\teg$ have
the same final state angular distributions, then
this is incompatible with hierarchical
couplings in $\SX$. So  one more
condition  is required to test a hierarchy
in the $\SX$:\\
$~$\\
4) $~$ in decays of polarised $\tau$s, the angular 
distribution of
final state leptons can be measured \\

However, New Physics which induces $\SL$
and $\SR$ may also induce $\T_X$, as was
the case for the leptoquark $S_0$. 
Consider   a hierarchical  $\T$,
neglecting at first the $\SX$.  
The $\llg$ branching ratios will therefore
be parametrised by $t_3$, and the
third rows of $V_L$ and $V_R$. This makes five
parameters (as opposed to three for $\SL$ or $\SR$),
so  the polarised  branching ratios are required
to hope to test the hierarchical pattern
in $\T$. Suppose, for instance, that $\tau_L \to
\mu_R \gamma$ is measured. Both  $\widetilde{BR}(\mu_L \to
e_R \gamma)$ and  $\widetilde{BR}(\mu_R \to
e_L \gamma)$  are small, so, in particular
\beq
\label{Brrtau2}
\frac{\widetilde{BR}(\mu_R \to e_L \gamma)}
{\widetilde{BR}(\tau_L \to \mu_R \gamma)}
=\frac{  |{ T}_{e  \mu }|^2}
{ |{ T}_{ \tau \mu}|^2}
=\frac{|V^*_{L3 e} V_{R3 \mu}|^2 }
{|V^*_{L 3 \tau}V_{R3 \mu}|^2 } \lsim 10^{-4}
\label{BRrtau2}
\eeq
so with  assumption 3) above,
$\widetilde{BR}(\tau_R \to e_L \gamma)
\propto |V^*_{L3 e} V_{R3 \tau}|^2$ should
be beyond the reach of Super-B factories.
Similarly, if  $\tau_R \to
e_L \gamma$ was measured, a hierarchical
structure in $\T$ predicts that
 $\tau_L \to
\mu_R \gamma$ is inaccessible.

  \begin{table}
\renewcommand{\arraystretch}{1.25}
$$
\begin{array}{|c||c c|c c|c c c c|}\hline 
\widetilde{BR} &\SL  &\SL &\SR &\SR  &\T  &\T   &\T  &\T  \\
\hline
\tau_R \to \mu_L \gamma 
& \checkmark & x 
& x & x 
&\checkmark &
&x & \\
\tau_L \to \mu_R \gamma 
&x   &x
& \checkmark & x
&  &\checkmark
&  & x\\
\tau_R \to e_L \gamma 
&x & \checkmark 
&x & x 
&  & x
&  &\checkmark \\
\tau_L \to e_R \gamma 
&x  &x 
&x &\checkmark
&x & 
&\checkmark &\\
\hline
\end{array}
$$
\caption{If the polarised decay indicated by $\checkmark$ is
observed, then a hierarchical pattern in the operators listed
in the top row predicts that the decays indicated by
$x$ should not be seen at Super-B factories. The blank
spaces mean that no prediction is made.  This 
table depends on the assumptions 1-4 listed in the 
text, see in particular assumption 3). 
The operators $\SL$ and $\SR$ can induce either of their two
colomns, $\T$ can induce any two non-conflicting colomns.
See the discussion at the end of section \ref{sec:hiertest}. 
\label{tab:2}}
\renewcommand{\arraystretch}{1.00}
\end{table}
Table \ref{tab:2} summarises  which processes
should not be seen, given a polarised
decay induced by a particular flavour tensor.
For instance, if  
one  $\tau  \to e_{X \b} \gamma$ decay is observed,
this is consistent with a hierarchical ${\bf S}_X$
or $\T$.
If both one of the  $\tau_R \to e_{L \b} \gamma$,
and  one of the  $\tau_L \to e_{R\b} \gamma$ are observed,
this is consistent with simultaneous
hierarchical $\SR$ and $\SL$. 
If the particular combinations,  $\tau_R \to \mu_L \gamma$
and  $\tau_R \to e_L \gamma$, or 
$\tau_L \to \mu_R \gamma$
and  $\tau_L \to e_R \gamma$,
are observed, this is consistent with
 a hierarchy in $\T$. So  hierarchies
in the various  flavour tensors  predict
that various rates are suppressed. However, 
if all the $\tlg$ decays are observed
in all polarisations, this is consistent
with the simultaneous presence of
hierarchical $\SX$ {\it and} $\T$.
That is,    a hierarchical pattern in  $\T$ 
or  in the  $\SX$ make opposite predictions:
if, for instance, $\tau_R \to \mu_L \gamma $
is observed, this is compatible with
${\bf S}_L$, in which a hierarchy imposes
that  $\tau_R \to e_L \gamma $ must not be seen.
However,  $\tau_R \to \mu_L \gamma $ could
also be induced by $\T$, where a hierarchy
is consistent with  $\tau_R \to e_L \gamma $,
but not  $\tau_L \to e_R \gamma $. 
  Therefore,
 with the  assumptions made so far, 
the hierarchical structure cannot be falsified. 
In addition, one must
assume that\\
$~$\\
5)   {\it either} the $\SX$,  {\it or} $\T$ is dominant. \\
$~$\\
Then it is clear from the table that a
 hierarchical structure
 can be falsified: for instance, if
$\tau_L \to \mu_L \gamma$  and
$\tau_L \to e_L \gamma$ are observed,
this is inconsistent with  a hierarchy
in $\SL$ (or $\SR$).

\subsection{Hierarchies and  invariants}
\label{sec:hieretinvar}

This subsection  makes contact between flavour tensors as invariants,
and the prospects of testing their  hierarchical structure. Section
\ref{sec:seesawhier}
discusses
the  conditions  which the  spurions
should  satisfy,
such as to obtain  a hierarchical invariant. 
Section \ref{sec:LQhier} uses invariants to study
whether the leptoquark $S_0$, which can
induce $\SX$ and $\T$, satisfies assumption five
for testing a hierarchy.

\subsubsection{Getting a hierarchy in $\SX$}
\label{sec:seesawhier}

Ideally,   the hierarchy condition 
 eqn (\ref{suffgrand})
could be translated to conditions which
the spurions should satisfy to
ensure a hierarchical invariant. However, this 
phenomenological
approach amounts to
expressing elegant invariants in
explicit basis-dependent form,
which defeats the purpose of invariants,
and risks to produce
opaque interlinked bounds on unknown
eigenvalues and mixing angles.
Instead, it is simple to  study
the issue in the wrong direction,
that is, start from
hierarchies in the spurions, and
enquire when then are transmitted
to the flavour tensors.

Two simple cases are: 
\ben
\item  degenerate eigenvalues  $M_I$ of the matrix $M$ of
the  New Particles \\
Consider to be concrete the
seesaw case. If $\lambda$ is written,  in the charged lepton and
heavy singlet mass eigenstate bases, as $V_L^\dagger D_\lambda V_R$,
then for degenerate singlet mass -squared $|M_I|^2$,
the hierarchy  condition on $S$
eqn (\ref{suffgrand})  is trivially applied to
$\lambda^\dagger \lambda$ in the charged lepton
mass eigenstate basis:
\beq
[V_L]_{j \b} \geq 
\frac{\lambda_i}{\lambda_j} [V_L]_{i \b} ~~~~ i \leq j
\label{sg1}
\eeq
To explore the area around  the degenerate mass limit,
one can write 
\bea
[\lambda D_M^{-2} \lambda^\dagger]_{\b \a} & =&
 \frac{1}{M_1^2}\left(  [\lambda \lambda^\dagger]_{\b \a}
-
[\lambda]_{\b J}  \frac{M_1^2 - M_J^2}{M^2_J} \lambda^*_{\a J} 
\right)
\eea
and assume that the first term satisfies
eqn (\ref{suffgrand}). 
Then
the second term is negligeable if 
\beq
M_J^2 - M_1^2 \ll M_J^2
\eeq

\item  flavour-independent  vertices \\  
If,  in general,  $\lambda \equiv V_L^\dagger  D_\lambda V_R$,
then for $D_\lambda$  proportional to
the identity matrix,  the hierarchy condition of
eqn (\ref{suffgrand}) applies to $V =V_L^\dagger  V_R$ and $D_M^{-2}$.  
\een

Illuminating formulae for the case where
 there
are hierarchies in $\lambda$ and $M$ are not
obvious to find. 

 For $\widetilde{S_0}$,  without generations indices,
 $\SR$  will
have hierarchical form if 
$ \widetilde{\lambda}_R^* \widetilde{\lambda}_R^T
\equiv V_e  D_{\tilde{\lambda}}^2 V_e^\dagger$ 
is hierarchical on its lepton indices (equivalently, 
if $V_{e}$  and  $D_{\tilde{\lambda}}^2$  satisfy conditions
(\ref{suffgrand})). 
If  $\widetilde{S}_0$  has
generation indices, then for three leptoquarks
of mass squared   $M_I^2$,  the same conditions
as discussed for the seesaw are sufficient to
obtain a hierarchical  flavour tensor.
Similar comments would
apply to the leptoquark $S_0$ with couplings $\lambda_L$ or
$\lambda_R$.

\subsubsection{$\T$ versus $\SX$ for the $S_0$ leptoquark}
 \label{sec:LQhier}

The fifth condition for  testing/confirming a hierarchy,
is that only one of  $\T$ or $\SX$ makes a significant
contribution to $\tlg$. 
The leptoquark $S_0$, 
with both  couplings $\lambda_L$ and
$\lambda_R$,  induces
both structures of matrix element,
so it is interesting to compare the relative 
size of $\T$ and $\SX$.

\begin{figure}[ht]
\unitlength.5mm
\begin{center}
\epsfig{file=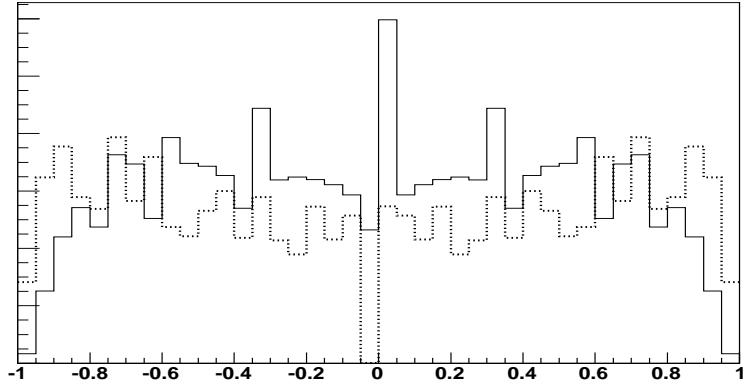,height=6cm,width=12cm}
\end{center}
\caption{This plot illustrates the importance of
 assumption 5) for testing hierarchies.  An
$S_0$ leptoquark coupled principally to $c$ quarks
would induce hierarchical flavour tensors, but can give
similar rates to the four final states
$e_L \gamma$, $e_R \gamma$
$\mu_L \gamma$, and  $\mu_R \gamma$. The
solid  and dotted lines are respectively
 the asymmetries $A_{e \mu}$ and
 $Ae_{LR}$ (see eqn (\ref{asyms})), obtained
after scanning over allowed couplings
(see text around eqn (\ref{asyms})).
\label{fig:sad} }
\end{figure}

First, notice that   constraints from two-quark-two-lepton 
contact interactions apply directly to   the magnitude of terms contributing
to the invariant. This is a useful
feature of invariants. 
The bounds  from 
$\tau \to e_\b \pi$ \cite{LQrev} impose 
that the $u$ quark contribution to
$[\SX]_{\b \tau}$ is less than $2 \times 10^{-4}$,
   so insufficient to
generate an observable $\tlg$ rate (see
eqn (\ref{voir})). On the other hand,
a diagram with
$c$ quarks in the loop can 
induce an observable $\tlg$ via $\SX$ or $\T$;
there are ${\cal O}(1)$ bounds on
$[\SR]_{\b \tau}$ fron
$Z \to \overline{e_\b} \tau$,  contraints 
from measuring
$V_{cs}$
on
$[\SL]_{\b \tau}$ which are
an order of magnitude smaller, and finally
bounds from $D_s \to \tau \nu_\b$ 
on the $c$-loop contribution to $\T$
 which  (just) allow it to contribute
an observable $\tlg$ rate.  Finally,
it is clear that  loops involving a top quark
can contribute to an observable rate
via at most two of $\SX$ and $\T$. This
is because one cannot simultaneously obtain 
$[\lambda_L]^*_{\tau t}
[\lambda_L]_{\b t}
  m_\tau \sim
[\lambda_R]^*_{\tau t}
[\lambda_R]_{\b t}
m_\tau
\sim
[\lambda_L]^*_{\tau t}
[\lambda_R]_{\b t}
m_t
\sim
[\lambda_R]^*_{\tau t}
[\lambda_L]_{\b t}
m_t$.
So  unfortunately,
$S_0$  leptquarks interacting principally with a $c$ quark
could have hierarchical couplings that generate similar
rates for the four processes $\tau \to e_{L,R} \gamma$ and
 $\tau \to \mu_{L,R} \gamma$. This is illustrated in
figure \ref{fig:sad},  which shows flat  asymmetries
$A_{e \mu}$(solid) and $Ae_{LR}$ (dotted), defined as
\beq
A_{e \mu } \equiv
\frac{\widetilde{BR}(\tmg) - \widetilde{BR}(\teg)}
{\widetilde{BR}(\tmg) + \widetilde{BR}(\teg)}
~~~~,~~~
Ae_{LR} \equiv
\frac{\widetilde{BR}(\teLg) - \widetilde{BR}(\teRg)}
{\widetilde{BR}(\teg)}
~~~.
\label{asyms}
\eeq
An asymmetry for muons $A\mu_{LR}$ can also
be defined; it has the same distribution
as  $Ae_{LR}$. 
The histograms are obtained by varying the six  couplings
$\lambda_{L \ell c}, \lambda_{R \ell c}$ (for 
$\ell \in \{ \tau, \mu , e \}$) between .01
and 1 with a log prior, and computing the asymmetries
for all the points which give a total rate
for $\tlg$ below the current bounds and within
the reach of SuperB Factories.  The normalisation
of the  vertical
axis is arbitrary.

\section{Discussion}
\label{sec:disc}

The observed neutrino masses demonstrate
the presence of New Physics in the lepton sector.
However, we do not know what it is. 
 One approach to this problem is
to calculate  in  motivated models,
thereby restricting their parameters and identifyng
their  predictions. Alternatively, one
can try to ``reconstruct'' the New Physics
from data. This  paper aims to follow the second,
more phenomenological philosophy.

\subsection{Invariants}
\label{sec:discinvar}

If the masses and couplings constants of 
the new particles were known,  the
 dipole ``vertex functions'' ${\bf X}_{L,R}$ 
of eqn (\ref{dipole}), which are observable,
could be calculated to arbitrary accuracy.
However,  since we know neither the
new particle identities, nor their mass scale,
nor the structure of their flavoured
couplings, it would be convenient to
be able to separate
these unknowns   in  ${\bf X}_{L,R}$. 
So the first aim of this paper was to
study, for a variety of new particle
contents, whether it was possible to
factorize  ${\bf X}_{L,R}$ as  the product
of a dimensionful, unflavoured constant,
multiplying an  ``invariant''. 
We refer to flavoured coupling matrices
from the Lagrangian as ``spurions'', and define
invariants to be products of spurions.   Since
 ${\bf X}_{L,R}$ is a dimensionful matrix in flavour
spaces, it  can always be written
as a dimensionful factor multiplying a dimensionless
tensor in flavour spaces, which we
call a ``flavour tensor'' (see eqn (\ref{Xmfv}). The crucial question
is whether  this flavour tensor is an invariant.
This is an interesting question for phenomenology,
because it is comparatively simple to
obtain information about spurions (the Lagrangian
flavour structures) from invariants.

In this paper, we only consider one loop
coontributions in New Physics
models without a conserved parity. This means
that in the loop, there is always a Standard Model
particle, and a new particle. In such models, it seems 
that the question about invariants
 reduces to whether
the   ${\bf X}_{L,R}$ are dominated by
the coefficient of the dimension six operator
of eqn (\ref{v1}).  This claim is
based on comparing the formulae for
the  ${\bf X}_{L,R}$, to the Effective
Field Theory expansion in $1/\Lambda_{NPFC}$,
where  $\Lambda_{NPFC}$ is the scale of
new flavour changing physics in the lepton sector.
In eqn (\ref{Xmfv}), the   ${\bf X}_{L,R}$,
of mass dimension -1, 
are written as a sum of dimensionful
constants $\sim 4G_F  v/(16 \pi^2 \sqrt{2})$,
 multiplying  dimensionless flavour tensors.
The flavour tensors relevant to $\llg$,
are ${\bf S}_{L,R}$ and  ${\bf T}_{L,R}$,
which respectively enccde flavour change
among leptons of the same SU(2) representation,
or chirality-changing flavour change.
 Formulae for these  flavour
tensors are   in eqns
(\ref{Lav1}) to (\ref{Lav5}). They involve
dimensionless $k$ and $y$ functions of the mass ratio in the loop,
which are given in the  Appendix.
Assuming that the new particle masses are
at the scale $\Lambda_{NPFC} > v$,,  these $k$ and $y$ functions
can be expanded in $v^2/\Lambda_{NPFC}^2$, because
one of the masses is  a Standard Modele mass.
The coefficients in this expansion will
be combinations of coupling constants.
The terms in this expansion
can be identified  to  an Effective Field
Theory \cite{Grev,GEFT} expansion in $1/\Lambda_{NPFC}$, of
operators which can contribute to the
dipole interaction. These include the usual
dipole  operator
( eqn (\ref{v1}) and first term below)
at ${\cal O}(1/\Lambda_{NPFC}^2)$,
plus gauge invariant operators with an
arbitrary number of extra Higgs fields:
\beq
\frac{{\bf C}_{\b \a}^{(6)}}{\Lambda_{NPFC}^2}
(\overline{\ell}_\b H) \sigma^{\mu \nu} P_R e_\a F_{\mu \nu}
+ 
\frac{{\bf C}_{\b \a}^{(8)}}{\Lambda_{NPFC}^4} H^\dagger H
(\overline{\ell}_\b H) \sigma^{\mu \nu} P_R e_\a F_{\mu \nu}
+ ...
\label{disc1}
\eeq
In this way of  thinking, ${\bf C}^{(6)}$ is our invariant,
and it will be a good approximation to the dipole
vertex functions if  higher order terms
in eqn (\ref{disc1}) can be neglected, 
as would be natural for 
$\Lambda_{NP} \gg v$. 
However,  the New Physics scale
should be $\lsim 10$ TeV \footnote{
Normally, for such  ``TeV-scale'' New Physics, 
the coefficient ${\bf C}^{(6)}$ is matched
to the dipole vertex function at the scale
where the new particles are removed from the
theory. 
} to see
$\tlg$ at a Super-B Factory.

 Two models were considered, to
explore the  possibility 
that  the coefficient matrix   ${\bf C}^{(6)}$
is  an invariant,  and an adequate
approximation to the dipole vertex function.
In the case of a non-SUSY, TeV-scale type I seesaw,
for which the Lagrangian is given in
eqn (\ref{L}),
the new physics interacts with the doublet
leptons, so contributes
to $\SL$.
Two types of singlet scalar
leptoquark were also considered,  respectively coupled to down 
or up-type quarks.
If a generation number is attributed to
the leptoquarks, then the coupling
constants at the vertices on either side
of the loop are three index tensors, and
the invariant is obtained
by summing  generation  on  both 
internal lines. 
 At ${\cal O}(1/\Lambda^2_{NPFC})$,
the flavour tensors $\SX$  due to
leptoquarks or the seesaw are 
proportional to the invariant 
$$\lambda_2 {\bf M}^{-2} \lambda_1^\dagger
$$
(see eqns (\ref{SRseesaw}),(\ref{SRLAMSt0}),
  and (\ref{quisuisje})), 
where ${\bf M}^2$ is the mass-squared
matrix of the new particles, and
$\lambda_{1}$ and $\lambda_{2}$   are the
new couplings appearing  in the left and
right side of the blobs on the top row of
figure \ref{fig:ST} (In the models
considered in this paper, $\lambda_1 = \lambda_2$).

The leptoquark $S_0$ could also  interact with 
singlet and doublet fermions via the coupling
matrices $\lambda_R$ and $\lambda_L$.
The chirality flip neccessary for the  dipole
can arise  on the internal quark line, allowing
$S_0$ to generate the flavour tensor $\T$ represented
in the second row of figure \ref{fig:ST}. This
gives invariants of the form
$$
 \lambda_L {\bf Y}_u   {\bf M}^{-2} \lambda^{\dagger}_R
~~~,~~~
 \lambda_R {\bf Y}^{\dagger}_u   M^{-2} \lambda^{\dagger}_L
$$
(see eqn (\ref{TLAMS0})).
From these cases, we can extrapolate that the flavour structure
of  ${\bf C}^{(6)}$
is obtained 
by multiplying one SM Yukawa matrix with
two new flavoured couplings matrices, and
possibly an inverse mass-squared matrix
for flavoured new particles
(a dimensionless invariant can 
still  be obtained by normalising by
the smallest flavoured new particle mass$^2$ $M_1^2$).
 The form of
the invariant can be obtained by 
inspecting the relevant Feynman diagrams: 
vertices give a coupling constant, and
heavy propagators give ${\bf M}^{-2}$.

It remains the question  of whether 
the vertex functions ${\bf X}_{L,R}$ 
can reasonably be replaced by the 
 coefficient  ${\bf C}^{(6)}$ of
dimension six operators, which is proportional
to an invariant.  As discussed
after eqn (\ref{disc1}), there are corrections 
proportional to $v^2/\Lambda_{NPFC}^2$,
which is $<1$ by assumption. The perturbative
expansion is therefore well defined. However,
in flavour physics, there can be small mixing angles
and hierarchical couplings, so the
relevant question is whether  the
${\cal O}(v^2/\Lambda_{NPFC}^2)$ terms
we wish to drop 
are small compared to the invariant that was retained.
The answer is model-dependent. In the case
of the leptoquark $\widetilde{S}_0$, 
the ${\cal O}(v^2/\Lambda_{NPFC}^2)$ terms
are multiplied by a down-type Yukawa coupling
squared, and it is reasonable to neglect
corrections $\lsim m_b^2/M_{\widetilde{S}_0}^2$.
However, in the seesaw model, the 
 ${\cal O}(v^2/\Lambda_{NPFC}^2)$ corrections
are $\sim m_W^2/M^2$  (for $M$ a TeV-scale singlet
neutrino mass), and for the leptoquark
$S_0$, they  are  $\sim m_t^2/M_{S_0}^2$. 
Such corrections could contribute an
observable $\tlg$ rate for  new particle
masses $\lsim$  2 TeV
(see eqns (\ref{BRS}) and (\ref{BRT})).

There can also arise logarithms of flavoured
mass matrices, as for example in eqn (\ref{TLAMS0}).
These could induce errors in the  reconstruction
of a  flavoured matrix $Y$, because  
  the eigenvalues of $Y$ and   $Y( 1 + \log Y)$
are different, even  though
they are diagonalised
by the same matrix.  This is
unfortunate. However, maybe
logarithmic cirrections can be
considered ``small'' in a first approximation.

Another stumbling block, in approximating
the dipole vertex function as an invariant
multiplied by an unknown constant, is that
 several  distinct invariants could contribute
to ${\bf C}^{(6)}$. This was the
case for the $S_0$ leptoquark, which could
induce  $\SX$, $\T_L$ and $\T_R$ (see eqns
(\ref{SRS0}) to (\ref{TLS0})).  If new spurions
are few, or if new particles are rare, 
this  problem may not arise. However,
as studied in section \ref{sec:hieretinvar},
the leptoquark $S_0$ can induce all
four flavour tensors with similar contributions
to $\tlg$, while remaining consistent with
phenomenological constraints. It could be
interesting to study  whether there is
a unique invariant 
in models, such as the supersymmetric seesaw, with
a single new spurion for the charged leptons
(the neutrino Yukawa matrix), but several new
particles.

Finally, the invariant   could be interpreted as a term
from an MFV expansion, as these terms  
are also constructed by multiplying flavoured
matrices from the Lagrangian (``spurions''). However,
there are many terms in an MFV expansion, 
multiplied by  arbitrary $\lsim 1$ constants. 
The aim  here is to obtain only  one
invariant, in the hope of reconstructing
the spurions, so  some justification  for rejecting more
complicated invariants is required. 
We suppose that sums over flavour indices
arise from loops, so are accompagnied by
a $1/(16 \pi^2)$, which could suppress
invariants containing more ``spurions''.
In addition, Yukawa matrices have a
Higgs leg, which 
 may be contracted in a loop,  which
naively  gives an additional  $1/(16 \pi^2)$
suppression. Alternatively, if the Higgs leg remains as a vev,  this
brings a factor of  $v/\Lambda_{NP}$.
In summary, considering  only
the simplest  invariants may be
justified, 
provided  that its reasonable to neglect   higher order
terms in  both the loop expansion, and the Effective
Field Theory expansion in $v/\Lambda_{NP}$.

\subsection{Testing a hierarchy?}
\label{sec:dischier}

The current bound   from $\meg$ of 
$$\frac{\Gamma(\meg)}{\Gamma(\mu \to e \nu \bar{\nu})}
\leq 2.4 \times 10^{-12}
$$  
is almost four orders of magnitude more restrictive
than the planned sensitivity of SuperB Factories to
$\teg$ and $\tmg$ (see table \ref{tab}). Pessimists
could interpret that the coefficients of
lepton flavour violating
operators  are suppressed,  so 
 $\teg$ and $\tmg$ will be beyond the reach of
Super B Factories.  
This paper takes the opposite
view that SuperB Factories will see
$\teg$ and/or $\tmg$. From a 
model-building perspective, this 
could arise because flavour change is
larger in the third generation. 
From a phenomenological perspective, 
one can only learn about New Physics by seeing it,
so to address  the question ``what  can the $\llg$ decays
 tell us about flavour
structure in the lepton sector?'',
 we  assume  that 
$\teg $ and/or $\tmg$ are observed. 
Furthermore, the smallest angle in
the CKM matrix is $V_{ub} \sim
4 \times 10^{-3}$, so a $10^{-4}$ suppression
due to a small mixing angle squared is ``standard''.

In practise, we study the less ambitious
question of whether a hierarchical
flavour tensor makes falsifiable predictions,
where the definition of hierarchical
is given in section  \ref{sec:notnhier}.
It is model-independent, and 
unrelated to the  invariants discussed
in the first part of the paper. 
We focus on  a hierarchical structure,
because
it has sufficiently few
parameters  to be predictive,
and because  the Yukawa matrices 
are hierachical.
A hierarchical flavour tensor
which induces $\teg$ or $\tmg$ at a Super-B Factory,
must contain a small parameter to
suppress $\meg$ below the current bound.
Section \ref{sec:hier}
presents the argument that
this small parameter 
also suppresses a
 $\tlg$ decay, implying
that a hierarchical flavour tensor
forbids the observation
of all the  modes  $\teLg$,
$\teRg$,  $\tmLg$ and $\tmRg$. This argument
rests upon five assumptions,
given in section \ref{sec:hiertest}. 
The third and fifth  appear problematic.
The third assumption is that new particle mass scale 
is $\gsim 600$ GeV, so that the
couplings which control the observed
$\tlg$ rate  are ``large''. This enforces
a hierarchy with respect to the couplings
which control $\meg$, and the other
$\tau \to e_\b' \gamma$ decay. The LHC
may verify this assumption in several models.
The fifth assumption is that the 
dipole vertex function should not have its
general form given in eqn (\ref{Xmfv}),
but rather should be dominated by 
either $\SX$ or $\T$.  This means that
New Physics should be  in one of the
doublet  or  singlet lepton
sectors (generating $\SX$), or,
if it  interacts with both, 
then the amplitudes with
chirality flip inside the loop ($\T$)
should dominate those with
chirality flip on the decaying fermion line ($\SX$).
This condition arises because,
for a given observed $\tlg$ process, 
$\SX$ and $\T$ predict different
 rates to be suppressed, as is summarised
in table \ref{tab:2}. An example
where this condition is not satisfied
is given in section \ref{sec:LQhier}.
However, if new particles  below
10 TeV are rare, this fifth assumption is
not unreasonable.

Finally,  if
 a hierarchy were measured, then it
 could be  interesting to suppose 
that the flavour tensor is an invariant,
and explore what this implies about the
flavoured matrices of the Lagrangian. 
Section \ref{sec:seesawhier} briefly
discusses the simpler question
of when  a hierarchy in the
Lagrangian flavour structures
is transmitted to the  flavour
tensors contributing to
 the dipole vertex function.

\subsection{Summary}

 Beyond the Standard Model physics  
is required  in the lepton sector for neutrino
masses.  If present below 10 TeV,
it  could  induce  observable
 $\llg$ decays  in upcoming experiments.
Observing such decays could give
information on the flavour structure of
this New Physics.

A hierarchical  (in flavour space) dipole vertex function
  (where hierarchical means dominated by its largest eigenvalue),
could be confirmed if some, but not
all of the $\tlg$ decays are observed. 
Were it possible to distinguish internal
from external line 
chirality flip, such  a hierarchy could be tested.

In simple  models (with few loop diagrams containing one SM particle),
the coefficient of the ${\cal O}(1/M_{BSM}^2)$ term
in the dipole vertex function is a ``Jarlskog-like invariant'',
meaning that it is obtained by multiplying flavoured matrices
from  the Lagrangian. It has the form
$\lambda_1M_{BSM}^{-2} \lambda^\dagger_2$, where
$\lambda_1$ and $\lambda_2$ are the flavoured
couplings on the two sides of the loop.
The invariant is  a better approximation to the
vertex function when the particles in the loop are
(very) weakly coupled to the Higgs. 
Invariants elegantly  give a linear relation 
to flavoured matrices in the Lagrangian, avoid
confusion about basis, and   may 
simplify the combination of bounds 
from various processes
on New Physics.

\subsection*{Acknowledgements}

I thank Sebastien Descotes-Genon for 
clarifying discussions, without which
this paper would not have been written, Paolo
Gambino for  illuminating explanations of
Effective Field Theory,  Gino
Isidori for several interesting  conversations, 
and the LPTHE group at Jussieu for a seminar invitation,
a warm welcome, and useful comments.


\section*{Appendix: Summary of results from Lavoura \cite{lavoura}}
\label{App:Lavoura}

Lavoura defines coefficients $\sigma_L, \sigma_R$, related as
$$
\frac{e}{16 \pi^2} X_L  = \frac{ i \sigma_L}{2}
$$
to the ${\bf X}_X$ used here. The $\sigma_X$
are written in terms of functions $k_i$, $k_f$, etc,
which are $m_i \times$ the definitions given
here ( $m_\a$ is already scaled out of
the definition of  $\SX$ in 
eqn (\ref{Xmfv})):
\bea
k(t) & = &  \frac{t^2 -5t -2}{12 (t-1)^3} + \frac{t \ln t}{2 (t-1)^4 }
\label{k1}\\
& \to & \frac{1}{6} +  \frac{t \ln t}{2}~~~,~~~ t \to 0 \nonumber \\
& \to & \frac{1}{12 t}  ~~~,~~~ t \to \infty \nonumber \\
\overline{k} (t) 
& = &  \frac{2t^2 +5t -1}{12 (t-1)^3} - \frac{t^2 \ln t}{2 (t-1)^4 }
\label{k1b}
\\
& \to & \frac{1}{12}  ~~~,~~~ t \to 0 \nonumber \\
& \to & \frac{1}{6 t}  ~~~,~~~ t \to \infty \nonumber \\
k_f(t) & = &  \frac{t -3}{2 (t-1)^2} + \frac{ \ln t}{ (t-1)^3 }
\label{k3}
\\
& \to & - \frac{3}{2} -   \ln t~~~,~~~ t \to 0 \nonumber \\
& \to & \frac{1}{2 t}  ~~~,~~~ t \to \infty \nonumber \\
\overline{k_f}(t) 
& = &  \frac{t +1}{2 (t-1)^2} - \frac{t \ln t}{ (t-1)^3 }
\label{k3b}
\\
& \to & \frac{1}{2} + t \ln t  ~~~,~~~ t \to 0 \nonumber \\
& \to & \frac{1}{2 t}  ~~~,~~~ t \to \infty \nonumber \\ 
\label{y1bar}
\overline{y}(t) & =& 
 \frac{-4t^3 + 45t^2 -33t +10}{12 (t-1)^3}
- \frac{ 3t^3 \ln t}{2 (t-1)^4}
\\
& \to & -\frac{1}{3}  
{\Big (}1 - \frac{33}{4t}  + \frac{9 \ln t}{2t} ... 
~~~~~~t \to \infty \nonumber \\
& \to & -\frac{5}{6}  
{\Big ( } 1 
-\frac{3 t}{10}  + ...
~~~~~~t \to 0 \nonumber 
\eea
where $t = m_f^2/m_B^2$,  and  $m_B$ is  the boson mass.

\end{document}